\documentclass{aa} 

\usepackage{natbib}                                  
\usepackage{ulem}
\usepackage{graphicx}
\usepackage{txfonts}
\begin{document}
\title{The very low mass multiple system LHS\,1070 -- a testbed for model atmospheres for the lower end of the main sequence}

\titlerunning{Spectral analysis of the very low mass multiple system LHS\,1070}
\authorrunning{A.~S.\ Rajpurohit~et al.}

   \author{A. S. Rajpurohit\inst{1}, C. Reyl\'{e}\inst{1}, M. Schultheis\inst{1}, Ch. Leinert\inst{2}, F. Allard\inst{3},  D. Homeier\inst{3}, T. Ratzka\inst{4}, P. Abraham\inst{4}, B. Moster\inst{2,7}, S. Witte\inst{5}, N. Ryde\inst{6}
          }
   \institute{Universit\'e de Franche Comt\'{e}, Institut UTINAM CNRS 6213, Observatoire des Sciences de l'Univers THETA de Franche-Comt\'{e}, Observatoire de Besan\c{c}on, BP 1615, 25010 Besan\c{c}on Cedex, France
	  \and
	     Max-Planck-Institut f\"{u}r Astronomie, K\"{o}nigstuhl 17, 69117 Heidelberg, Germany       
	  \and
	    CRAL, UMR 5574, CNRS, Universit\'e de Lyon,  \'Ecole Normale Sup\'erieure de Lyon,  46 All\'ee d'Italie, F-69364 Lyon Cedex 07, France
        	  \and
	    Universit\"ats-Sternwarte M\"unchen, Ludwig-Maximilians-Universit\"at, Scheinerstr. 1, 81679 M\"unchen, Germany
	  \and
	  Hamburger Sternwarte, Gojenbergsweg 112, 21029 Hamburg, Germany
	  \and
	     Department of Astronomy and Theoretical Physics, Box 43, SE-221 00 Lund, Sweden
          \and
	     Max-Planck-Institut f\"{u}r Astrophysik, Karl-Schwarzschild-Str.1,
            85748 Garching, Germany
            }

 
 \abstract
{LHS\,1070 is a nearby multiple system of low mass stars. It is an
  important source of information for probing the low mass end of the
  main sequence, down to the hydrogen-burning limit. The primary of
  the system is a mid-M dwarf and two components are late-M to early L
  dwarfs, at the star-brown dwarf transition. Hence LHS\,1070 is a
  valuable object to understand the onset of dust formation in cool
  stellar atmospheres.} 
{This work aims at determining the fundamental stellar parameters of
  LHS\,1070 and to test recent model atmospheres: \mbox{BT-Dusty},
  \mbox{BT-Settl}, DRIFT, and MARCS models.} 
{Unlike in previous studies, we have performed a
  $\chi{^2}$-minimization comparing well calibrated optical and
  infrared spectra with recent cool star synthetic spectra leading to
  the determination of the physical stellar parameters
  $T_\mathrm{eff}$, radius, and $\mathrm{log}\,g$ for each of the
  three components of LHS\,1070. } 
{With exception of the MARCS models which do not include dust
  formation, the models are able to reproduce the observations and
  describe the main features of the visible to IR spectra. This is
  consistent with the fact that dust formation prevails in the B and C
  component atmospheres.  The parameters obtained with the DRIFT
  models confirm the values determined in earlier studies.   But
  important differences between models are observed, where the MARCS
  model is too bright in the $H$ and $K$ bands, and the BT-Settl and
  BT-Dusty models systematically yield up to 100\,K higher
  $T_\mathrm{eff}$ in the case of the B and C components. This
  confirms a trend for models without, or with less efficient cloud 
  formation, to predict higher $T_\mathrm{eff}$ than models richer in
  dust (DRIFT). Even models including cloud physics however still 
  produce slightly too bright $J$ band flux, showing as too blue $J-K$
  colors. The onset of dust formation remains therefore a
  particularly challenging regime to understand.} 
   {}


   \maketitle
%
\section{Introduction}
The lower end of the Hertzsprung-Russel diagram has much importance as the vast
majority of stars in the Galaxy are low mass stars. In the Galaxy,
70\% of the stars are M dwarfs. They contribute over 40\% of the total
stellar mass content \citep{Gould1996,Mera1996,Henry1998}. These
M dwarfs have a mass that ranges from $0.6 M_\odot$ to the hydrogen
burning limit of about 0.075 to $0.085 M_\odot$ depending on the metallicity
\citep{Chabrier2000b}. These stars are found in any population, from young
metal rich M-dwarfs in open clusters \citep{Reid1993,Leggett1994} to
the several billion years old metal poor dwarfs in the galactic halo
(Green and Morgan 1994) and in the globular clusters
\citep{Cool1996,Renzini1996}. Such low mass stars are an important
probe for our Galaxy as they carry fundamental information regarding
the stellar physics, galactic structure and formation, and its
dynamics. In addition, the existence of brown dwarfs or planets being
discovered and confirmed around M-dwarfs \citep[][and references
therein]{Butler2004,Bonfils2011} plays an important role in understanding the formation of brown dwarfs and planets.

Despite their large number in the Galaxy, little is known about low
mass stars because of the difficulty (i) to get a homogeneous sample
with respect to the age and metallicity due to their intrinsic
faintness, and (ii) to disentangle the parameter space
($T_\mathrm{eff}$, $\mathrm{log}\,g$, and metallicity). Indeed a
number of studies have shown that a change in temperature or gravity
can compensate for a change in metallicity to some degree. 
An additional difficulty is the complexity of their atmospheres:
convection in optically thin regimes, molecules, and dust cloud
formation for the later types. 
Water vapor and CO bands dominate the Rayleigh-Jeans branch of the
spectral energy distribution at infrared wavelengths ($> 1.3\, \mu$m),
while TiO, VO, and metal hydrides govern the corresponding visual ($>
4000\AA$) to near-infrared ($< 1.3\, \mu$m) spectral energy
distribution.  Convection reaches out to the optically thin (as far as to an optical depth 
$\tau \sim 10^{-3}$) portion of the atmospheres, flattening the
temperature gradient of the atmosphere \citep[]{Allard1997}. \cite{Ludwig2002,Ludwig2006}
have determined the mixing length based on a comparison of the mixing
length theory used in 1D static model atmospheres and Radiation
Hydrodynamic simulations. 
In M dwarfs later than M6 the outermost temperatures fall below the
condensation temperatures of silicate grains, which leads to the
formation of dust clouds \citep[see e.g.][]{Tsuji1996a,Tsuji1996b,
Allard1997,Ruiz1997,Allard1998a,Allard1998b}. These
processes complicate the understanding of these cool atmospheres. \\ 
One approach to study the physics at the low end of the main sequence
is to compare observed spectra with synthetic spectra from various
authors and modelling techniques. The determination of the physical
parameters (effective temperature, gravity, metallicity, radius) is
obtained by spectral synthesis, i.e. $\chi^2$ minimization. 

LHS\,1070 is a low mass multiple system of cool dwarfs discovered by
\cite{Leinert1994}, with visual magnitude 15. It is located at a
distance of 7.72$\pm$0.15 pc from the Sun \citep{Costa2005} and is
considered as a member of the disk population with a probable age
of around 1 Gyr \citep{Reiners2007a}. The spectral types for the A, B,
and C components were found to be M5.5-M6, M8.5, and M9-M9.5,
respectively \citep{leinert2000}. A fourth component was suspected
very close to the primary by \cite{Henry1999} from HST Fine Guidance
observations, but this detection is no longer considered to be real
(T. Henry, private communication).  
The latest orbit determination has been performed by \cite{K2012},
with semi major axes of 0.458'' for the close pair
 BC and 1.112'' for the wide orbit of BC around component A.
 E.g., on December 12, 2003,
 component B was separated from A by 1.77" at position 13$^{\circ}$,
 and component C from B by 0.41" at 178$^{\circ}$.

\citet{Leinert1998,leinert2000} have derived effective temperatures of
2950\,K, 2400\,K and 2300\,K for the components based on
spectral analysis, and found that B and C showed clear signatures of
dust in their spectra. They presented photometric mass estimates
ranging from 0.109 to $0.079 M_\odot$ for the three stars based on
theoretical isochrones, thus reaching right down to the minimum
hydrogen burning mass. This mass range makes LHS\,1070 a valuable
system for understanding the formation of dust in cool 
atmospheres and the processes that occur at the star/brown dwarf
transition. LHS\,1070 is therefore a testbed to validate and define
further developments of both atmospheric and interior models at  the
lower end of the main sequence.  We assume the same age and
composition for the three components of this system for simplicity. 

In this paper, we present the spectral synthesis of the components A,
B, and C. We determine their physical parameters by
comparing the well-calibrated HST spectra in the optical (from FOS)
and in near and mid-IR (from HST/NICMOS and ISOPHOT-S) with synthetic spectra
computed from recent stellar atmosphere models: BT-Dusty and BT-Settl
\citep[]{Allard2012a}, MARCS \citep{Gustafsson2008}, and
DRIFT \citep{Witte2009}. Observations and data reduction are described
in Sec.~\ref{obs}. Sec.~\ref{mod} presents the atmosphere models used
in the analysis. In Sec.~\ref{result} we give the determination of the
stellar parameters and show the comparison between observed and
modeled spectra. Discussion and conclusion follow  in
Sect.~\ref{ccl}. 


\section{Observations and data reduction}
\label{obs}
Regarding the optical photometry and spectroscopy, the reader is referred to \cite{leinert2000}. The new input concerns the infrared data.

\subsection{Photometry}

The J, H, K, and L' photometry presented in Table~1 for LHS\,1070 ABC refers to \cite{Leggett1998}. We obtained M-band photometry with the UIST instrument on UKIRT on
November 9, 2002, as well as N-band photometry with the 
MAX camera \citep{Robberto1998} on the same telescope on August 27, 1996.  The brightnesses of the individual components were then derived from the brightness ratios obtained with
NACO \citep{Rousset2003,Lenzen2003} for J, H, K on December 12, 2003, for
        L' and M on December 6, 2001. The MAX observations in N-band yielded separate brightnesses only for
component A and the sum B+C of the other two components. 

The M band data were obtained in the Mauna Kea Observatory Near-Infrared
    System ($\lambda_\mathrm{central}=4.7 \mu$m, 50\%-width$ = 0.23 \mu$m) in
    service mode. Aperture photometry was applied to the pipeline-reduced
    data. The N band photometry used a standard filter ($\lambda_\mathrm{central}
    =10.47 \mu$m, 5\%-width $= 4.65 \mu$m). After the standard processing
    steps (bad pixel correction, combination of individual
    chop cycles after shift-and-add) we performed aperture photometry.
    The absolute calibration at 10.4 $\mu$ relied on the HR~6464
    spectrophotometric standard model, produced by P. Hammersley using the procedure
    described in \cite{Hammersley.1998}, and made
    available on the ISO web page for ISO standards.
    HR~6464 has the same spectral type M0III as the standard HR~400
    actually observed, and the flux ratio of the two was determined
    from their fluxes in the IRAS 12 $\mu$m band. Strictly speaking, the  
    result is a narrow-band (0.25 $\mu$m) brightness
    at 10.4 $\mu$m under the assumption of an M0III type spectrum.
    The spectral slopes in
    this wavelength region are smooth and all are
    representing Rayleigh-Jeans tail emission. Uncertainties
    resulting from the difference in spectral type between M0III
    and our object therefore are not important.
    To be conservative, errors of $\pm$ 5 mJy, $\pm$ 4 mJy, and $\pm$ 3mJy
    are taken for the fluxes of the combined system, of component A,
    and the sum of components B and C, respectively.

Our NACO observations were used in
   determining the relative brightnesses of the three components
   of LHS\,1070, because of the superior spatial resolution of this
   instrument. After standard reduction (flat fielding, bad pixel
   correction), aperture photometry was applied.
   The NACO bandpasses closely match the MKO-NIR system for the
   J, H, K, L' bands. The main difference is in the N band (4.8 $\mu$m
   for NACO versus 4.7 $\mu$m at Mauna Kea, 20\% width of 0.64 $\mu$m
   for NACO versus 50\% width of 0.23 $\mu$m at Mauna Kea). In this wavelength
   range, the slopes of the component spectra are very similar.
   Therefore the NACO-measured brightness ratios were applied to extract
   the component brightnesses from the MKO-NIR based total brightnesses.

LHS\,1070 was observed with ISOCAM
\citep{Cesarsky1996} on the ISO satellite
in the LW2 (6.7 $\mu$m) and LW3 (14.3 $\mu$m) filters on November 28, 1996 (PI H. Zinnecker).
Near-infrared narrow and medium-band photometry from 0.90 $\mu$m to 2.15 $\mu$m was obtained with the NIC1 and NIC2 cameras of the HST NICMOS
instrument on Jan. 2, 1998. 
For the analysis we used pipeline reduced images.
      On each frame the system was clearly resolved, and it was possible to
      obtain separate photometric measurements for the three stars. Before
      performing aperture photometry for a given star, we removed the images
      of the other two objects by subtracting a scaled PSFPSF template,
      constructed from component
      A and  shifted to the known coordinates of the stars. An aperture
      correction, taken from the NICMOS Data Handbook, was applied, and
      photometric calibration was performed via multiplying by
      a conversion factor between the counts and F$_{\nu}$ as stated
      in the Handbook. The results are given in Table~1.

\begin{table*}
 \caption{Photometric data. Fluxes $F_\lambda$ are in log10(ergs
cm$^{-2}$ s$^{-1} \AA^{-1}$). 
}
\label{Table:1}
\begin{tabular}{c c c c c c c}
\hline
Wavelength  & Filter$^1$ & Component A  & Component B & Component C & 
\multicolumn{2}{c}{Components A+B+C} \\
$\mu$m & &\multicolumn{3}{c}{log10$(F_\lambda)$}&log10$(F_\lambda)$&original
notation \\
 \hline
 0.900& NIC1 F090M& $-13.347\pm0.010$&$-14.122\pm0.016$&$-14.289\pm0.017$ 
  && $156\pm2$ mJy$^2$ \\
 0.953& NIC1 F095N&$-13.213\pm0.031$&$-13.870\pm0.03$&$-14.056\pm0.036$  
  && $253\pm8$ mJy$^2$  \\
 0.970& NIC1 F097N&$-13.251\pm0.031$&$-13.943\pm0.027$&$-14.092\pm0.028$
  && $237\pm7$ mJy$^2$  \\
 1.083& NIC1 F108N&$-13.229\pm0.024$&$-13.831\pm0.017$&$-13.953\pm0.017$
  && $332\pm7$ mJy$^2$  \\
 1.100& NIC1 F110M&$-13.292\pm0.006$&$-13.887\pm0.007$&$-14.037\pm0.009$
  && $295\pm2$ mJy$^2$   \\
 1.130& NIC3 F113N&$-13.267\pm0.022$&$-13.809\pm0.073$&$-13.965\pm0.097$
  && $343\pm15$ mJy$^2$   \\
 1.130& NIC1 F113N&$-13.249\pm0.020$&$-13.796\pm0.019$&$-13.962\pm0.022$
  && $355\pm7$ mJy$^2$   \\
 1.25& J&$-13.350\pm0.017$&$-13.904\pm0.028$&$-14.036\pm0.028$&$-13.189\pm0.012$  & $9.14\pm0.03$ mag  \\
 1.450& NIC1 F145M$-13.494\pm0.005$&$-14.077\pm0.005$&$-14.218\pm0.005$
  && $326\pm2$ mJy$^2$  \\
 1.640& NIC1 F165M&$-13.517\pm0.004$&$-14.026\pm0.004$&$-14.163\pm0.004$
  && $419\pm2$ mJy$^2$  \\
 1.65&  H&$-13.535\pm-0.018$&$-14.061\pm0.028$&$-14.195\pm0.031$&$-13.367\pm0.013$&  $8.51\pm0.03$ mag \\
 1.660& NIC3 F166N&$-13.492\pm0.013$&$-13.930\pm0.084$&$-14.130\pm0.133$
  && $472\pm29$ mJy$^2$   \\
 1.800& NIC2 F180M&$-13.677\pm0.006$&$-14.197\pm0.008$&$-14.333\pm0.009$
  && $346\pm2$ mJy$^2$  \\
 1.900& NIC1 F190N&$-13.751\pm0.014$&$-14.276\pm0.007$&$-14.415\pm0.008$
  && $324\pm4$ mJy$^2$  \\
 2.040& NIC2 F204M&$-13.833\pm0.006$&$-14.343\pm0.004$&$-14.474\pm0.004$
  && $314\pm2$ mJy$^2$  \\
 2.150& NIC2 F215N&$-13.825\pm0.013$&$-14.273\pm0.005$&$-14.400\pm0.006$
  && $378\pm5$ mJy$^2$  \\
 2.2&  K&$-13.860\pm0.018$&$-14.339\pm0.018$&$-14.468\pm0.024$&$-13.674\pm0.014$  & $8.14\pm0.03$ mag \\
 2.300& NIC2 F222M&$-13.844\pm0.004$&$-14.301\pm0.003$&$-14.429\pm0.003$
  && $378\pm2$ mJy$^2$  \\
 2.375& NIC2 F237M&$-13.973\pm0.005$&$-14.447\pm0.003$&$-14.579\pm0.003$
  && $317\pm2$ mJy$^2$  \\
 3.8&  L'&$-14.604\pm0.013$&$-14.990\pm0.013$&$-15.097\pm0.014$&$-14.377\pm0.012$ & $7.63\pm0.06$ mag \\
 4.78&  M&$-15.008\pm0.019$&$-15.449\pm0.028$&$-15.536\pm0.041$&$-14.806\pm0.017$&  $7.72\pm0.04$ mag \\
 6.7  &LW2&                  &                 &                  &$-15.349\pm0.019$ &  $70\pm3$ mJy \\
 10.4&    N& $-16.190\pm0.053$ & \multicolumn{2}{c}{$-16.378\pm0.079$}
                  &$-15.973\pm0.035$&  $38.4\pm3$ mJy \\
 14.3&  LW3 &                 &                 &                  &$-16.555\pm0.043$&  $21\pm2$ mJy \\
\hline
\end{tabular}\\
$^1$ \,\, M: medium-band filter ($\Delta \lambda = 0.1-0.2 \mu$m), 
N: narrow-band filter ($\Delta \lambda = 0.02-0.04 \mu$m)\\
$^2$ \,\, these values are the sum of the values measured directly for 
the individual components
\label{tab:1}
\end{table*}

\subsection{Spectroscopic observations}
\label{obs-spec}

Near-infrared spectra of the individual components of LHS\,1070 were taken with grisms G096, G141 and G206 and the NIC3 camera of the 
HST NICMOS instrument on January 2, 1998 for the J, H, and K bands. 
In a first step we removed from the four spectra we had for
       each grism the short wavelength and long wavelength ends. For
       grism G096 this left the range of 0.80-1.085 $\mu$m or
       0.8-1.15 $\mu$m, depending on the local noise level.
       Similarly, G141 covered the range of 1.10-1.60 $\mu$m or
       1.10-1.85 $\mu$m. G206 ranged from 1.65 $\mu$m to 2.45 $\mu$m.
       For each wavelength pixel, all spectra  covering that
       wavelength were averaged using a weighted average, where
       the weights came from the formal uncertainties of the spectra (typically from 2\% to 5\%).
The reduction was based on the NICMOSlook IDL-based data 
reduction package. For the extraction of the spectra, the 'no weighting' option was used. The result for the 
A component was slightly scaled  (by a few percent) to fit the NICMOS photometry. The extraction procedure also gave the combined 
spectrum of components B and C (separated  only 0.4'' on the detector), which was decomposed into the spectra of the individual components on basis of the brightness 
ratio as a function of wavelength as determined from the narrow- and medium band photometry. The spectral resolution of the NICMOS spectra is 
$R = \lambda/\Delta\lambda= 200$.\\

\begin{figure*}[ht]
   \centering
   \includegraphics[width=6cm]{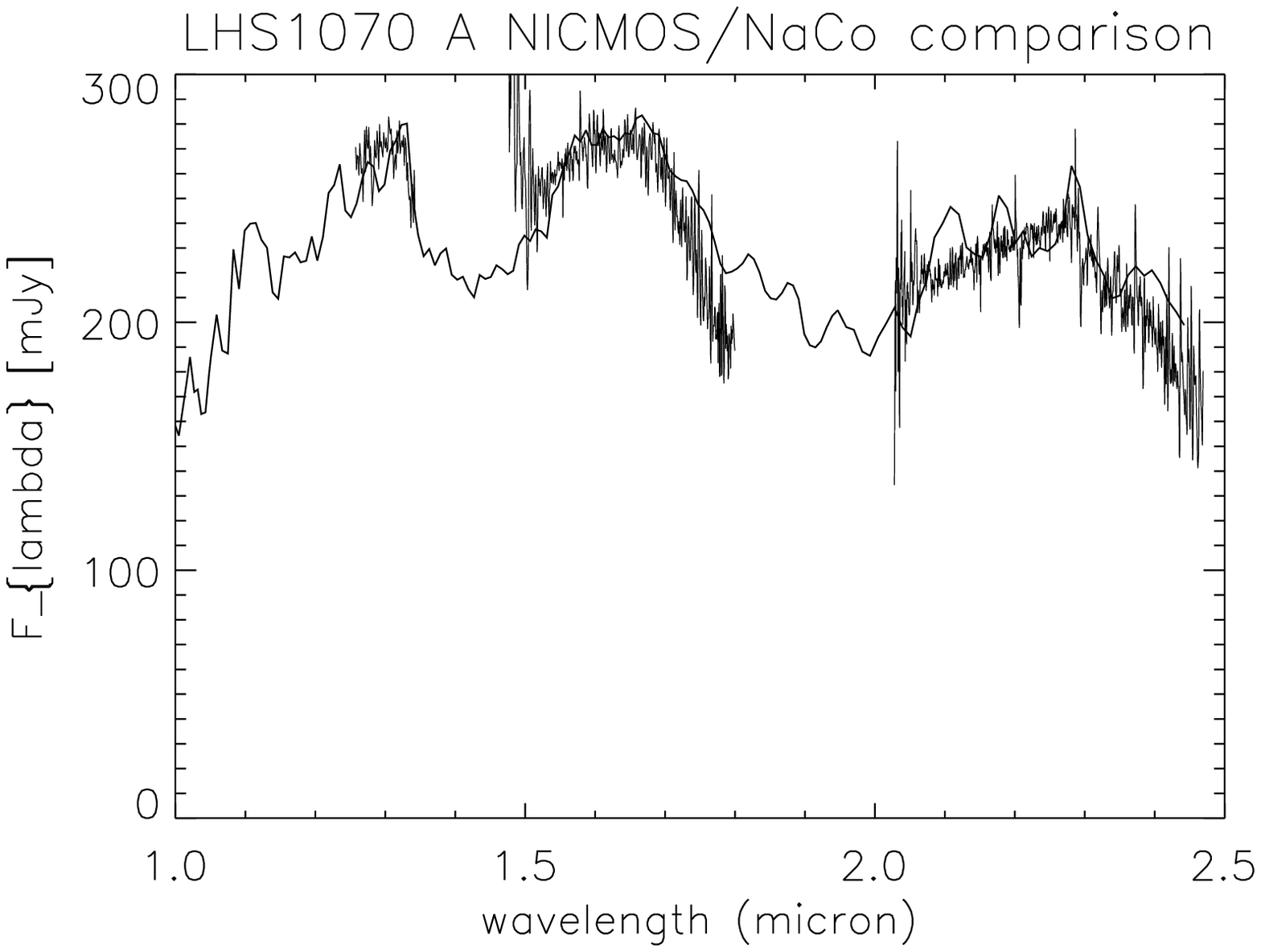}
   \includegraphics[width=6cm]{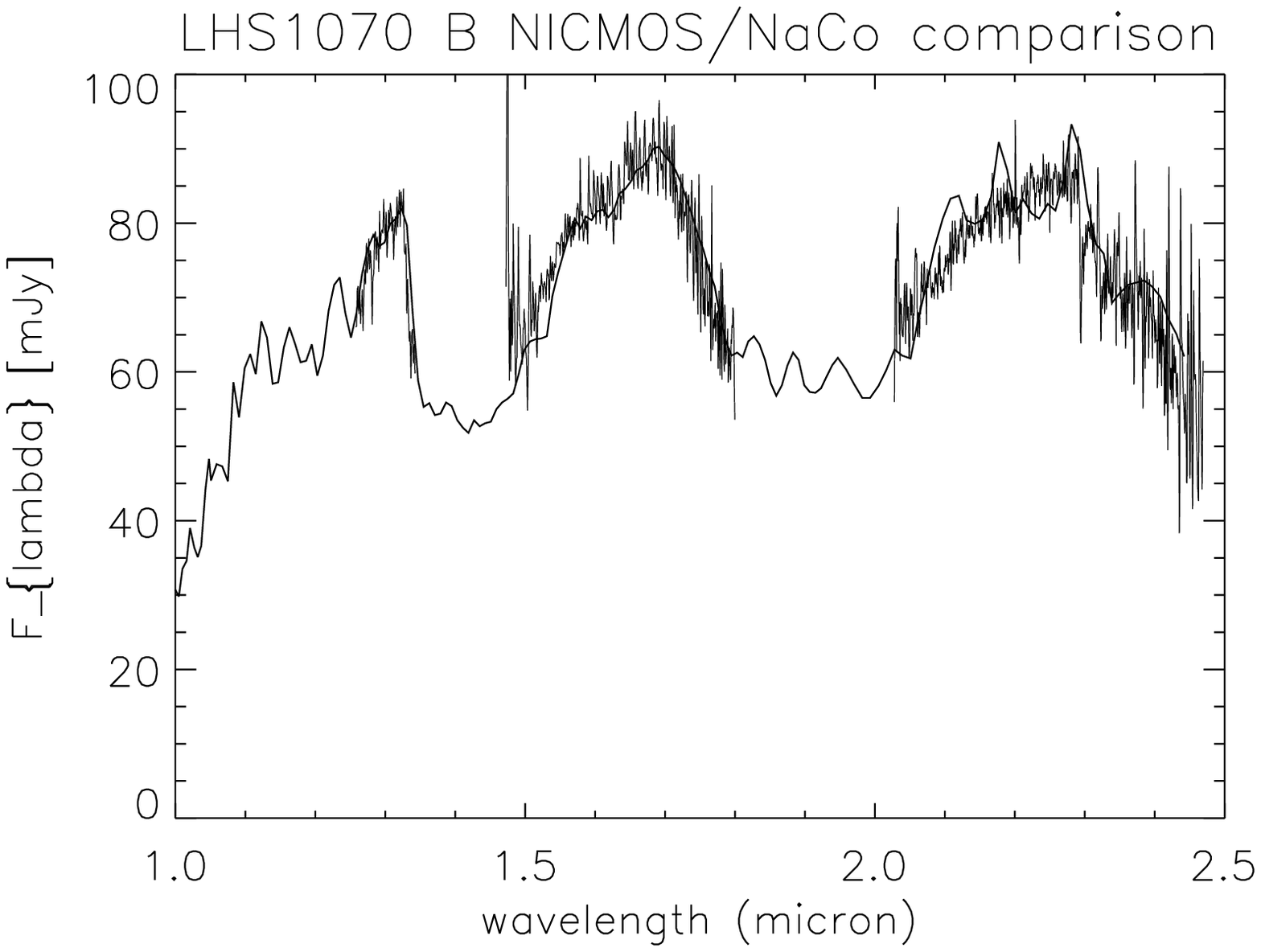}
   \includegraphics[width=6cm]{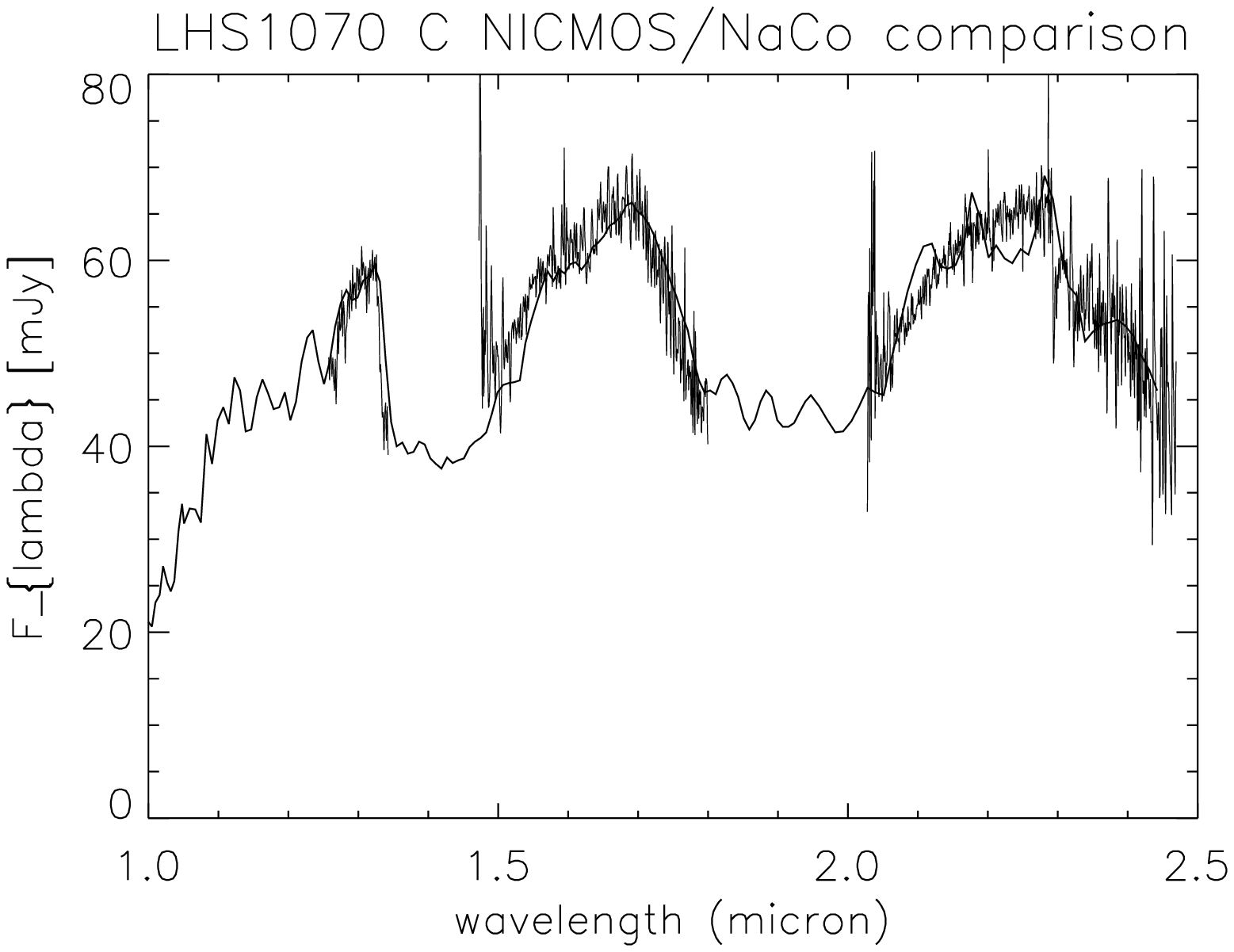}
      \caption{Comparison between NICMOS and NACO spectra of the three components. 
 }
         \label{Fig0}
   \end{figure*}

Higher resolution spectra with $R = 400, 1500, and 1400$ in the J, H, and K bands, respectively, were taken with the VLT NACO instrument
\citep{Rousset2003,Lenzen2003} on telescope UT4 of the VLT for the individual components of LHS\,1070. The camera/grism/filter settings were
S27/Grism4/J, S54/Grism3/H, and S27/Grism3/K, respectively. The slit width was 86 milli-arcsec in all three bands. After subtracting the sky
from the
raw data, a flat normalized along the dispersion direction was applied. The spectra were then traced and extracted. Finally, the individual exposures
were averaged. This procedure was also applied to the telluric standards. All telluric standards are of spectral type G2V, which allows a proper
modelling of their intrinsic spectrum with that of the Sun. The Solar spectrum was constructed using the data and scripts provided by
\cite{Maiolino1996}. Wavelength calibration was obtained by arc lamp exposures. The additional effort to guarantee the absolute flux calibration was
not taken, so only the relative shape of the spectrum was determined in each band. For each component, we used one scaling factor for each band to
bring the spectra to the absolute level as resulting from the NICMOS observations. The 
errors, taken as the formal uncertainties of the averaging process, typically 
range from 5-10\%. The number of pixels per resolution element in the J, H,
and K band is about 3, 1.5, and 3.
Because of the missing independent absolute calibration, the
        NACO spectra are used primarily for the study of spectral features.
        However, comparison with the NICMOS spectra, like in
        Figure~\ref{Fig0}, gives confidence also in their spectral shape.\\

Spectra for the system from 2.5 $\mu$m to 11.7 $\mu$m were obtained with the ISOPHOT-S spectrometer on board the ISO satellite on November 23, 1996
(PI: T. Tsuji) with an exposure time of 4096 s. ISOPHOT-S was the spectroscopic sub-instrument of ISOPHOT \citep{Lemke1996}. It had an entrance aperture of 24"x24", covering simultaneously the 2.5 to 4.9 and 5.9 to 11.7 $\mu$m ranges. The spectral
resolution of both channels ranged between 65 and 130. We processed the ISOPHOT-S observations using the {\tt Phot Interactive Analysis (PIA) V10.0}
\citep{Gabriel1997} following the standard data reduction scheme. The measurements were further reduced following our self-developed processing
scheme (K\'osp\'al et al. submitted), 
correcting for the slight off-centre positioning of the source.\\

\subsection{Spectroscopic features}

\begin{figure*}[ht]
   \centering
   \includegraphics[width=12cm]{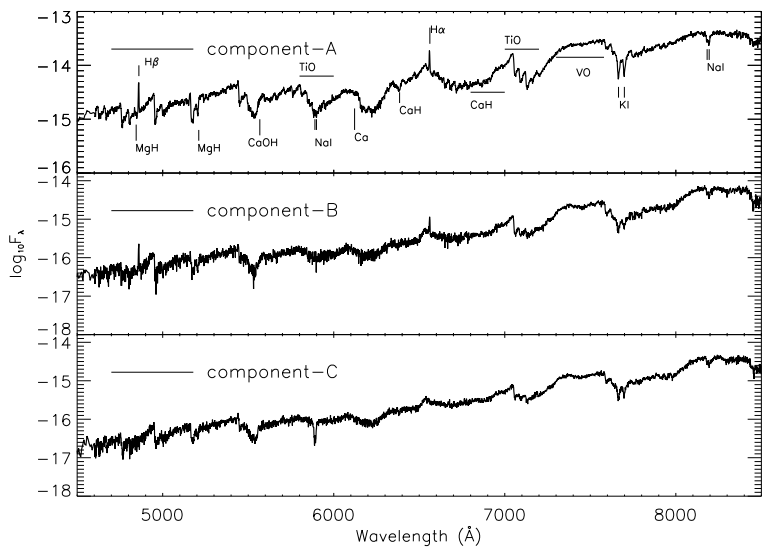}
      \caption{Optical Spectra of the components of LHS\,1070
          obtained with the Faint Object Spectrograph (FOS) on
          HST. The atomic and molecular features visible in all three
          components are shown in the upper panel. 
 }
         \label{Fig1}
   \end{figure*}

Figure~\ref{Fig1} shows the optical spectra of all the three components, with expected atomic and molecular features in the optical range between 5000
to 8500$\AA$. The most important ones are molecular bands of TiO, 
CaH, and VO, and atomic lines like CaI, NaI, and KI.
The TiO bands get weaker towards lower temperature, from component A to C, due to
condensation into dust species. The hydride CaH at 6380~$\AA$ and 6880~$\AA$ decreases in strength with decreasing temperature. The KI doublet at 7665 and 7699~$\AA$ is very strong and is useful for gravity determination. The NaI doublet at 8183~$\AA$ and 8195$\AA$ is strong in all the components, whereas CaI at 6103~$\AA$ is weak at all temperatures. \\

\begin{figure*}[ht]
   \centering
   \includegraphics[width=8cm]{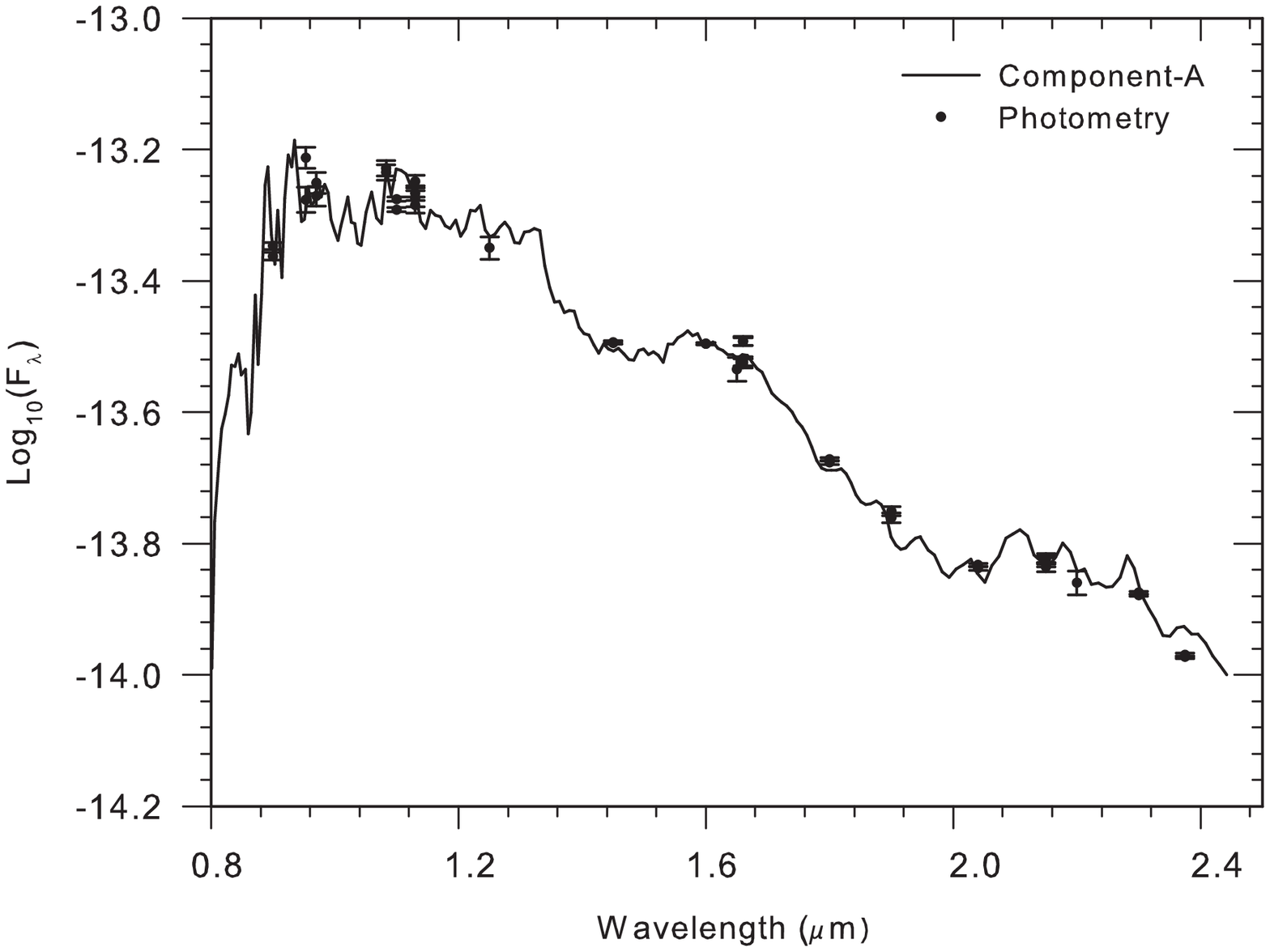}
   \includegraphics[width=8cm]{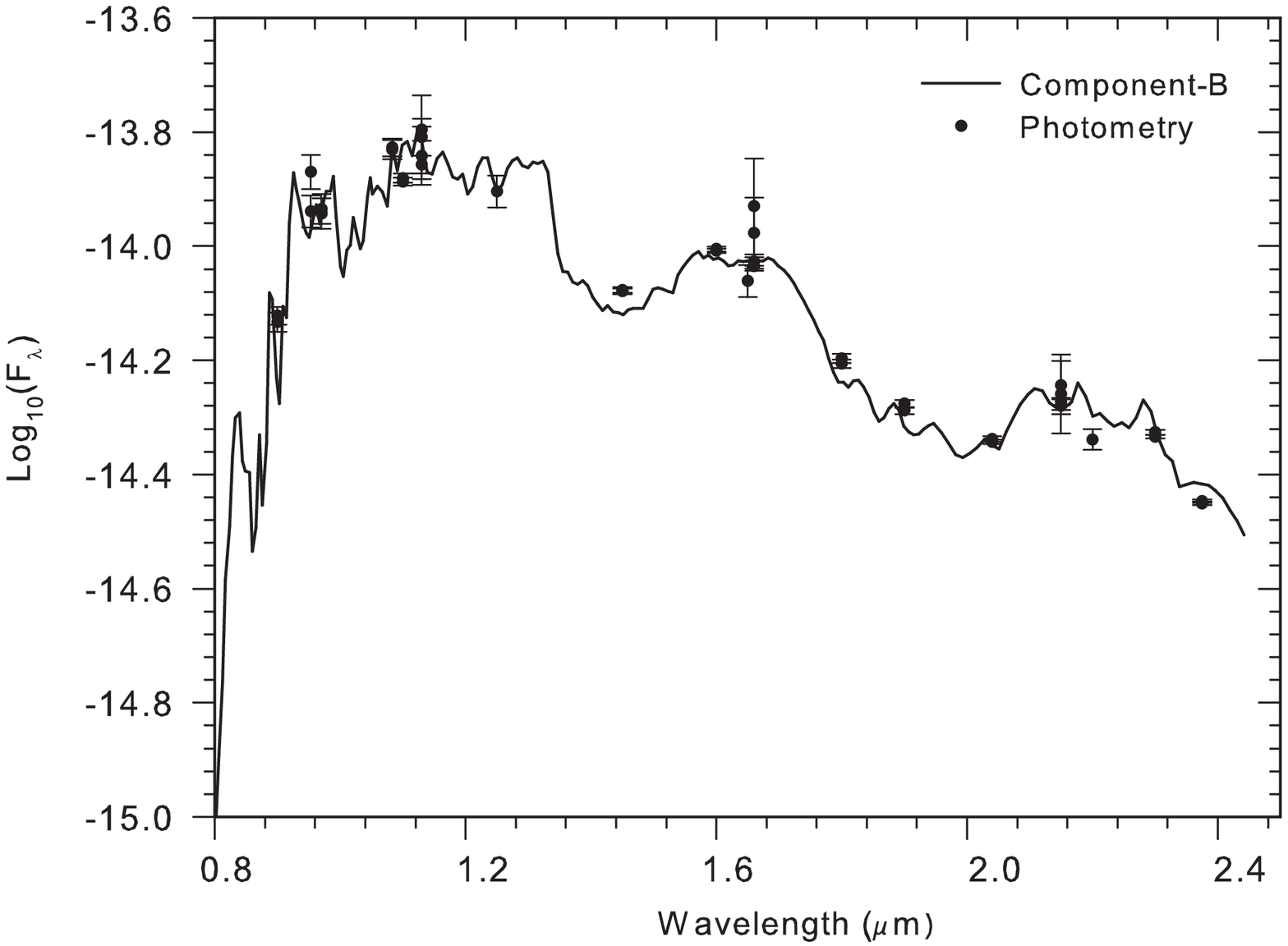}
   \includegraphics[width=8cm]{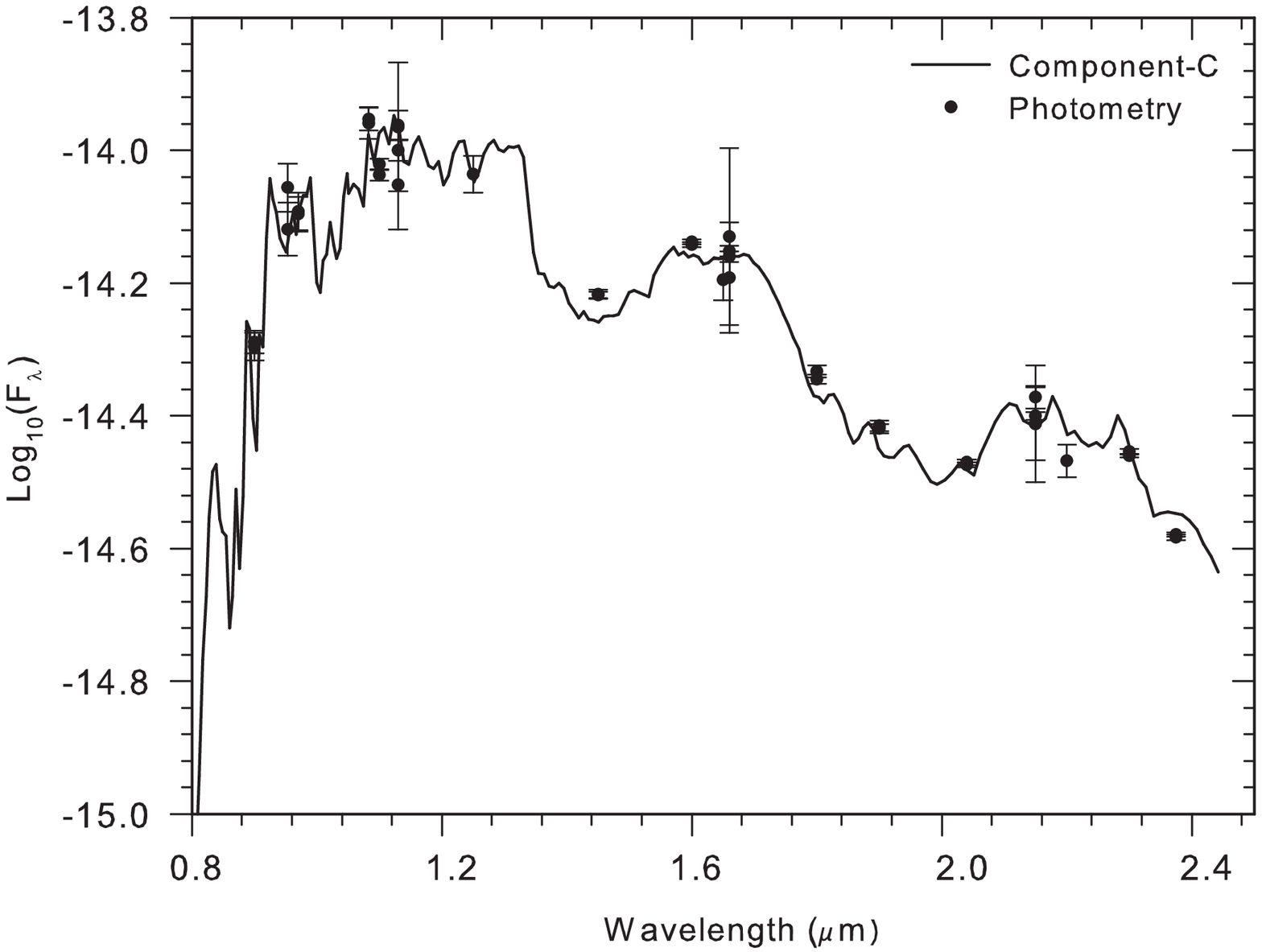}
    \includegraphics[width=8cm]{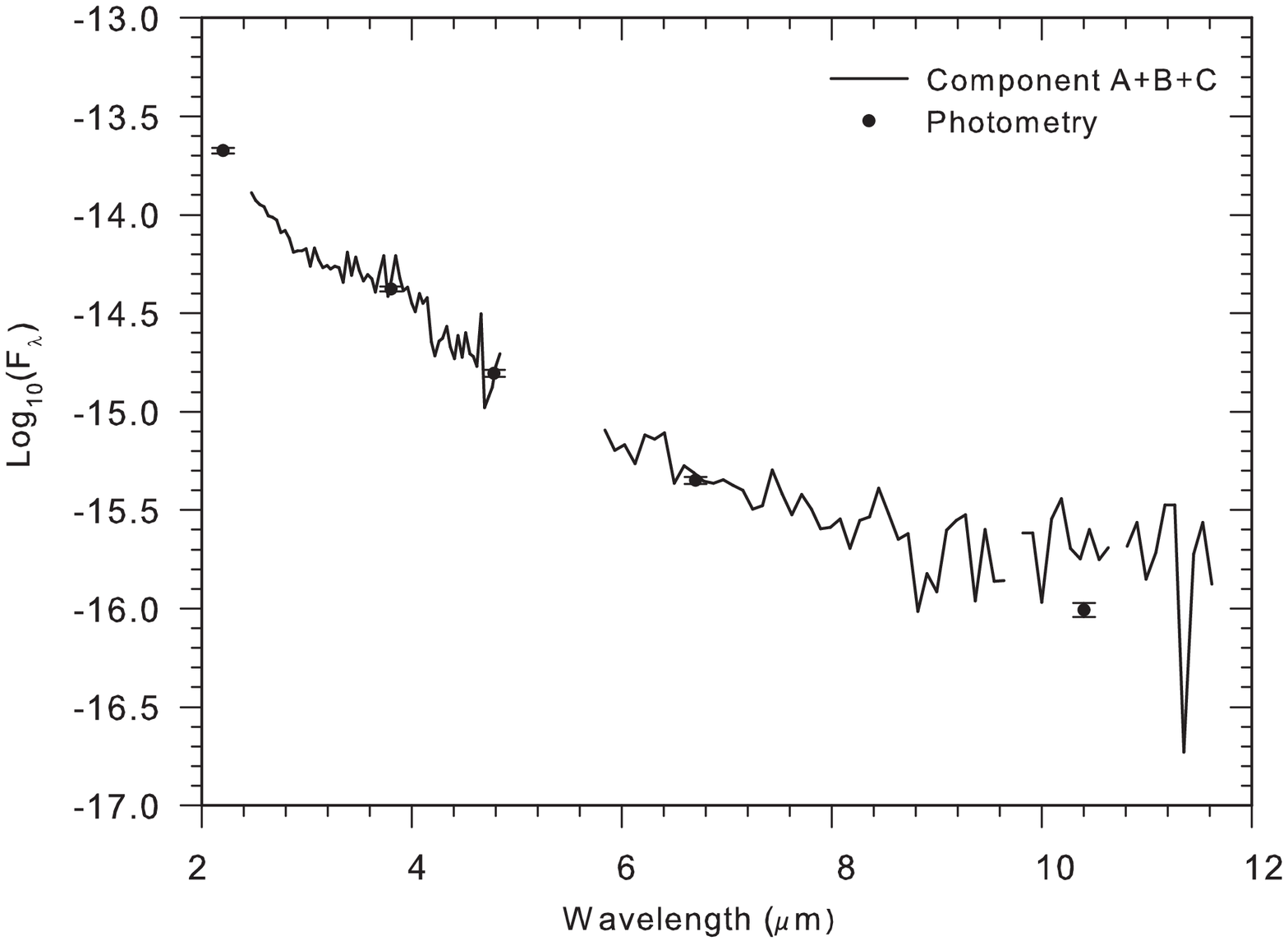}
      \caption{Near-IR spectra (solid line) and photometry (solid circles) obtained with NICMOS on HST for components A (upper left), B (upper right), and C (lower left). The ISOPHOT thermal infrared spectra of the unresolved system is shown on the lower right panel, again with photometric measurements 
overplotted.}
         \label{Fig2}
   \end{figure*}

 Figure\ref{Fig2} shows the near-IR spectra obtained with NICMOS on HST for 
 the three components as well as the thermal infrared spectrum of the unresolved system A+B+C taken with ISOPHOT on ISO. Photometric results are superimposed on these spectra.

\begin{figure*}[ht]
   \centering
   \includegraphics[width=6cm]{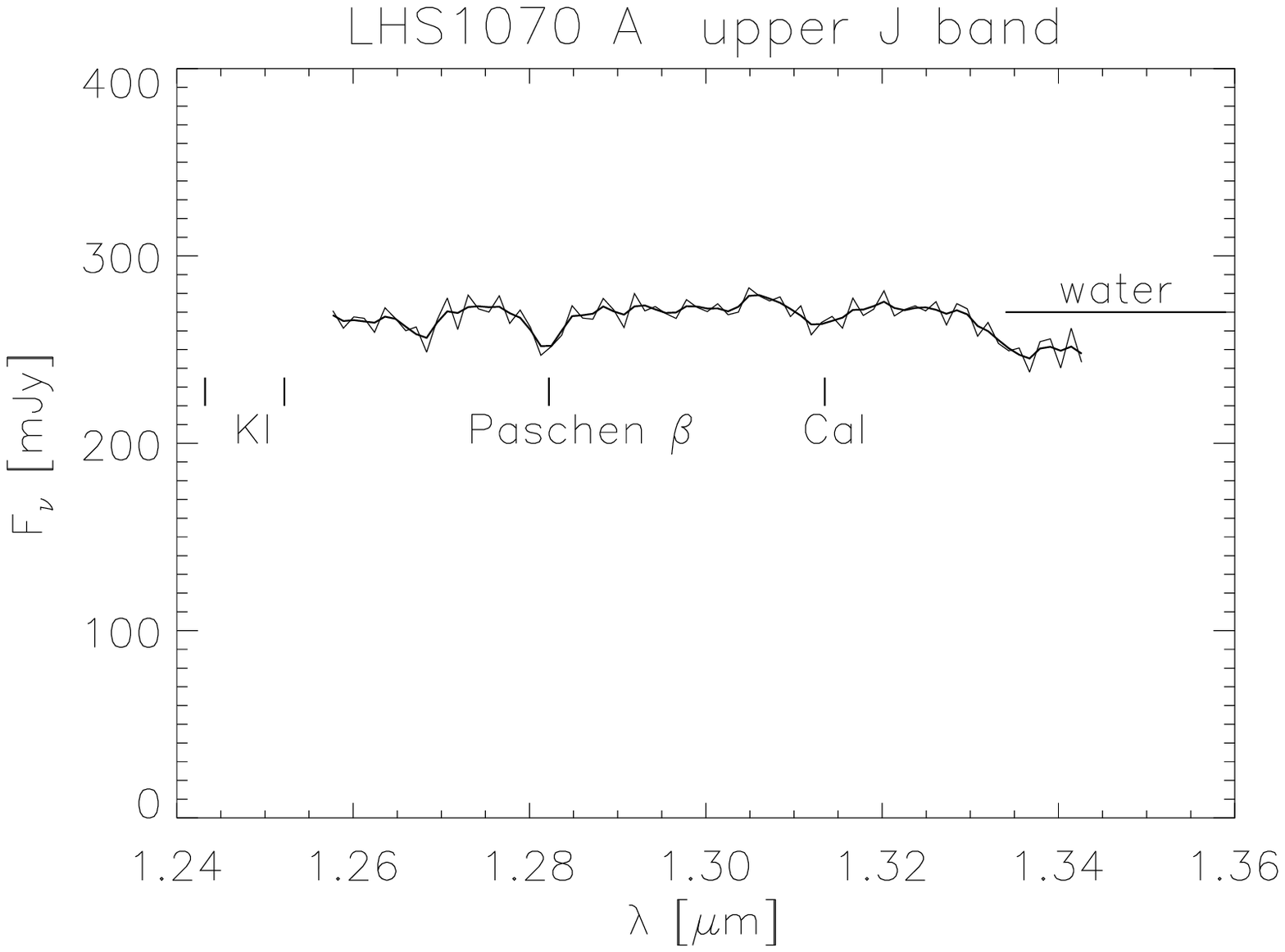}
   \includegraphics[width=6cm]{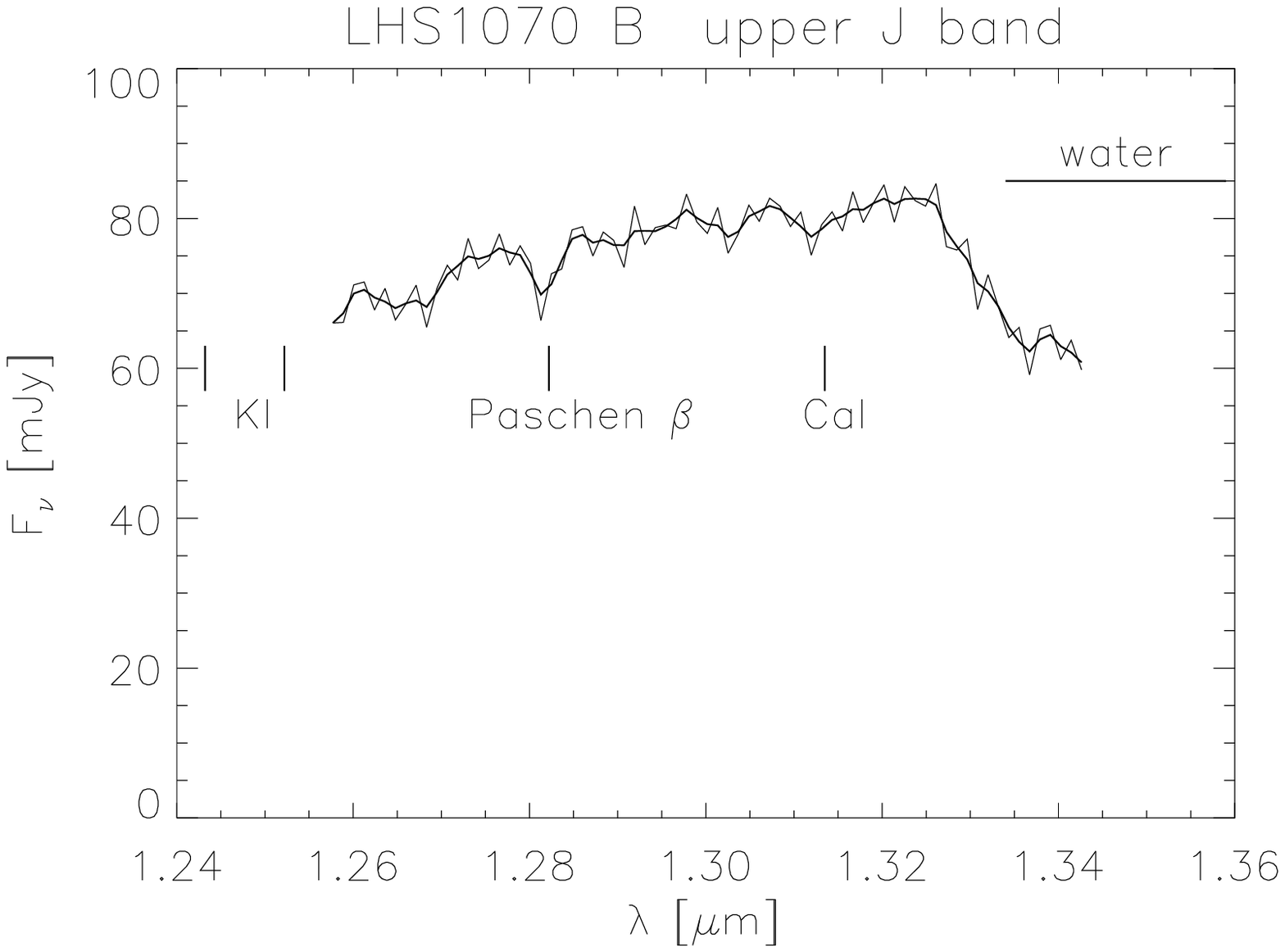}
   \includegraphics[width=6cm]{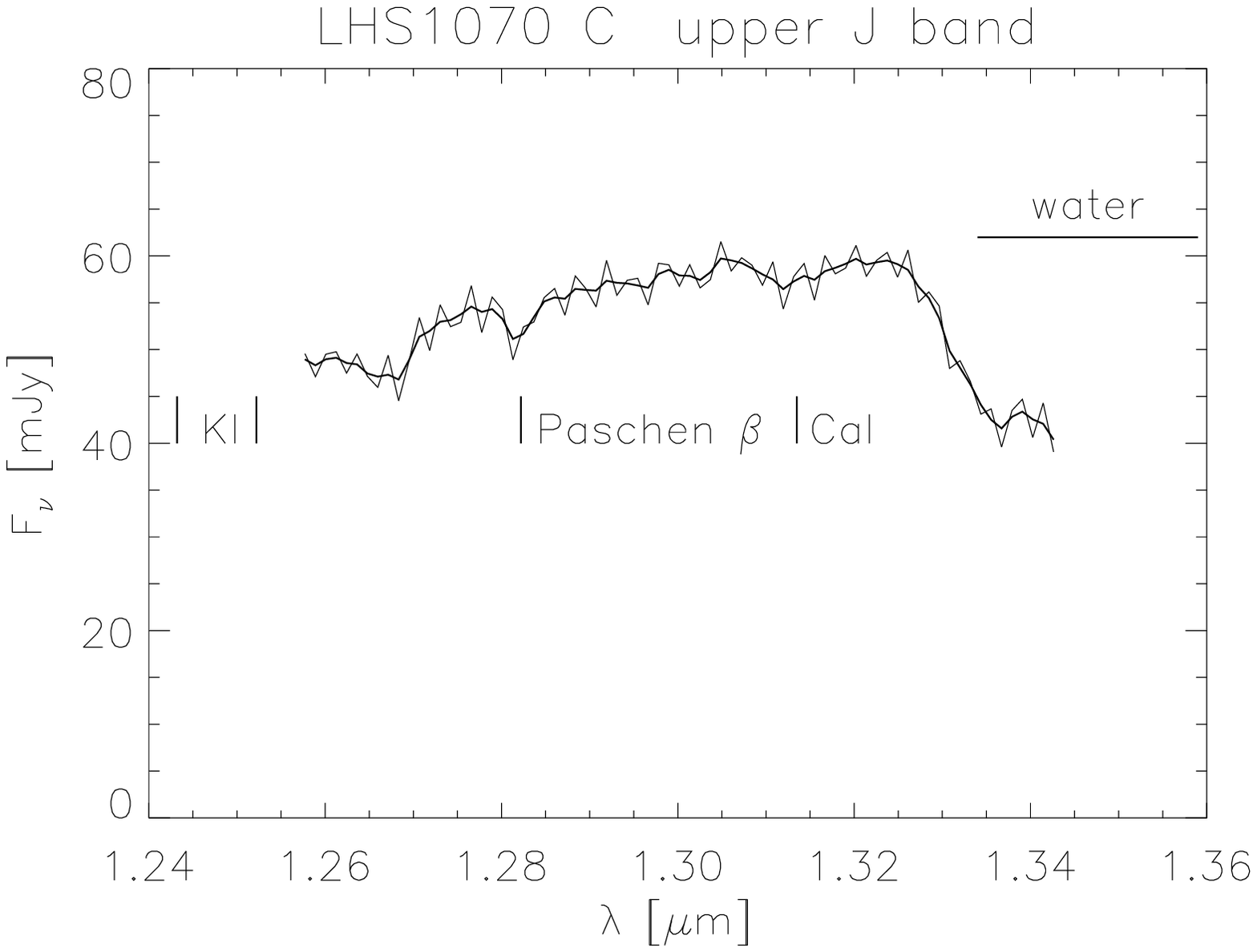}
   \includegraphics[width=6cm]{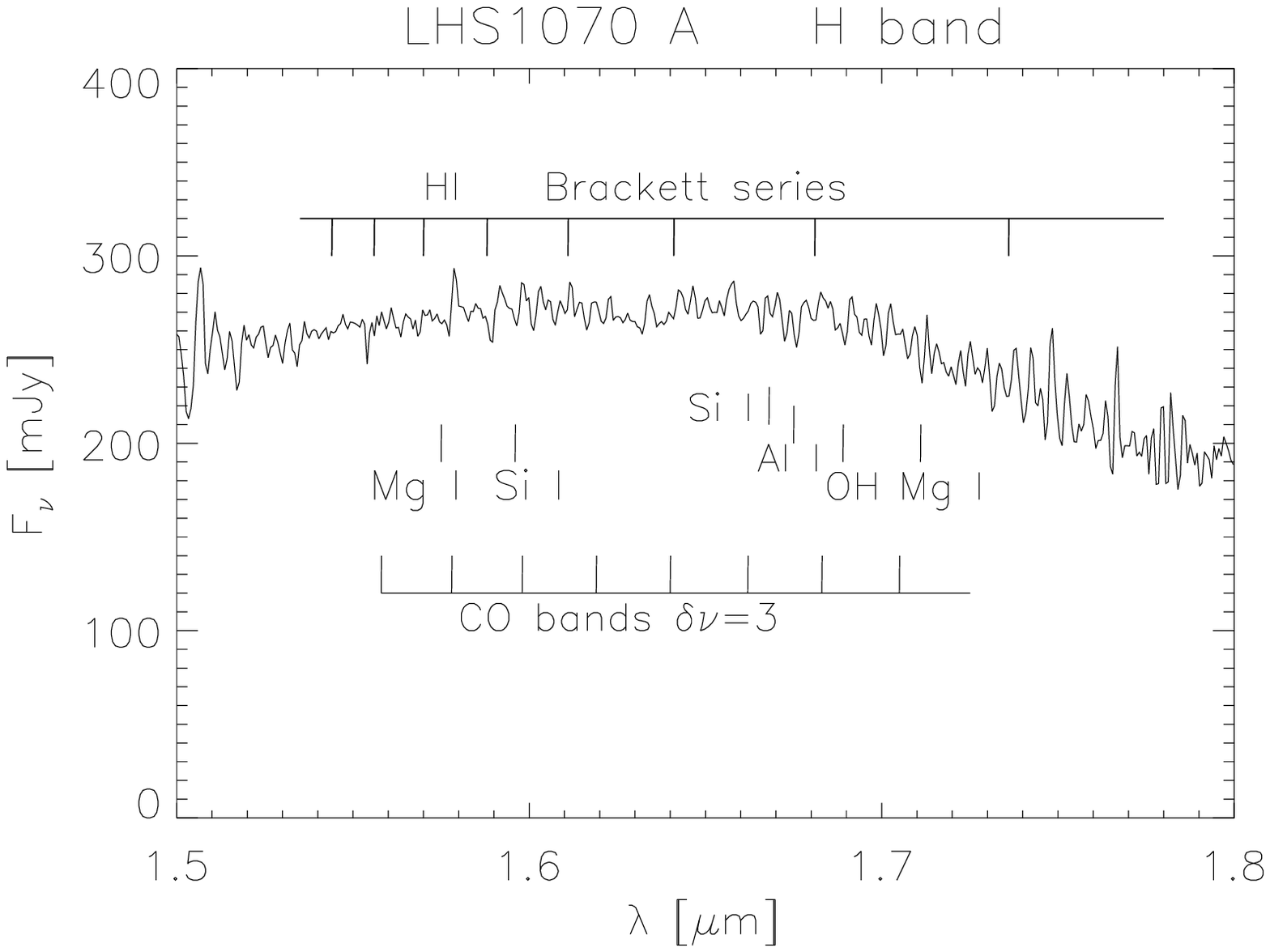}
   \includegraphics[width=6cm]{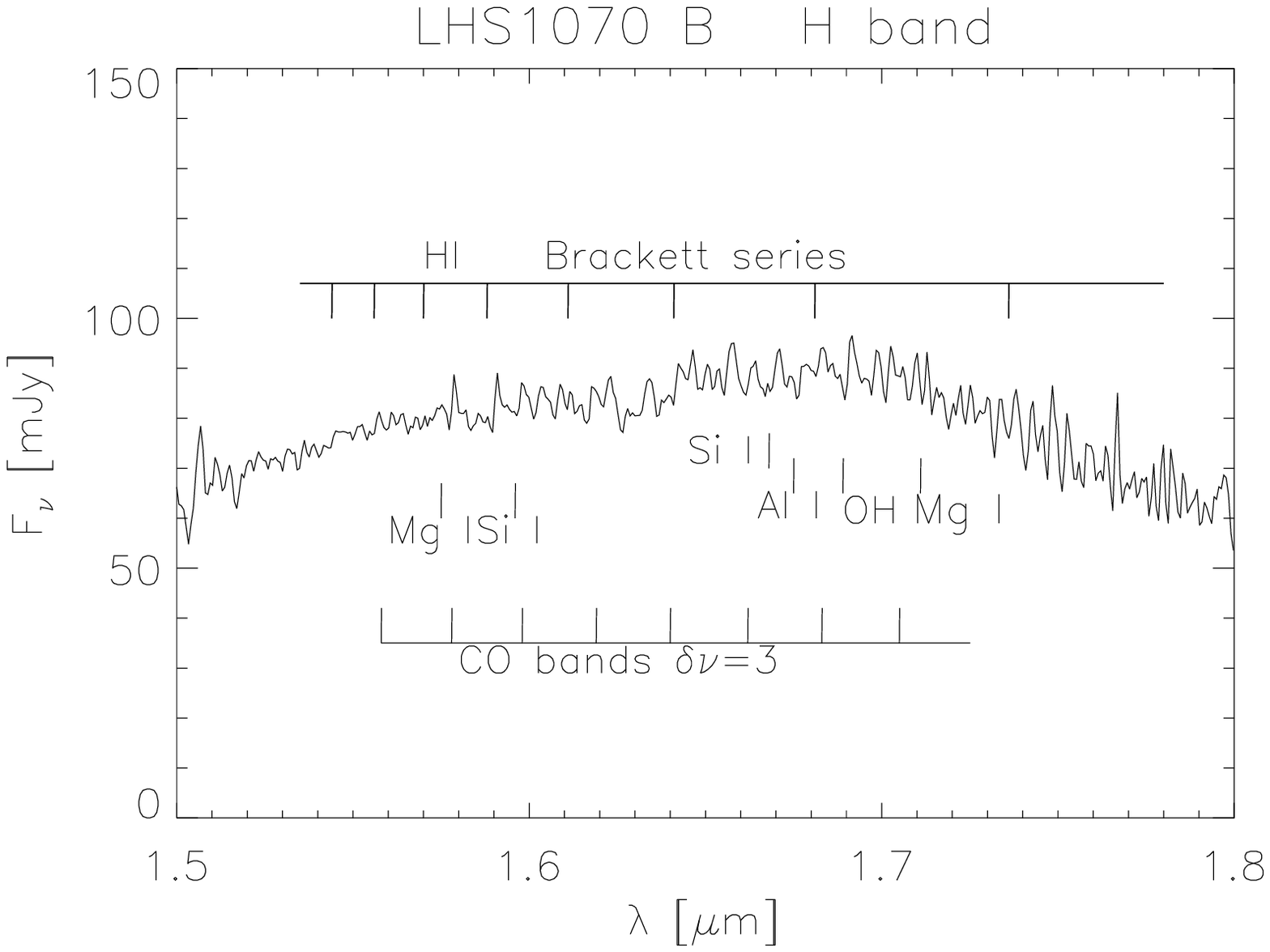}
   \includegraphics[width=6cm]{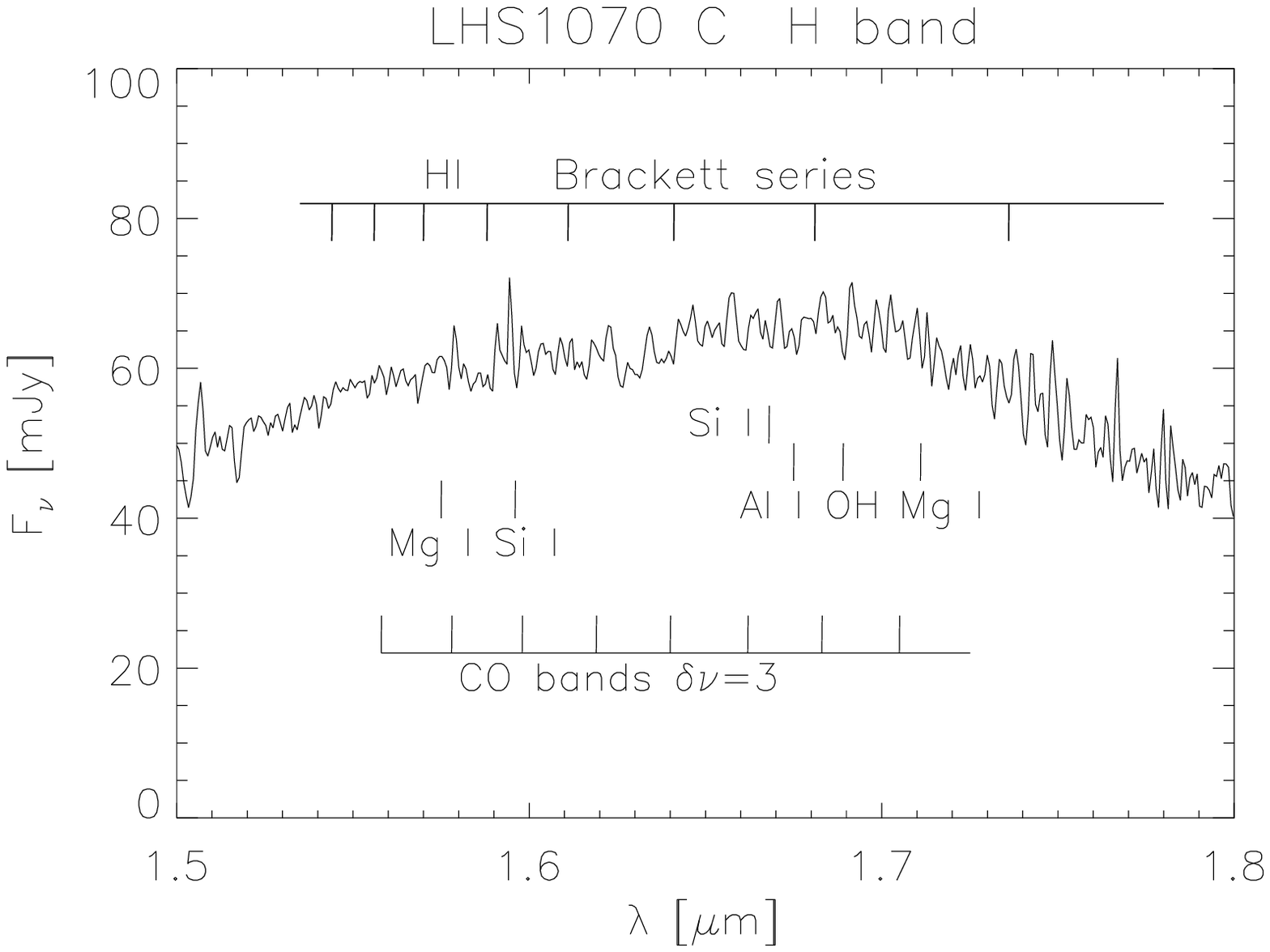}
   \includegraphics[width=6cm]{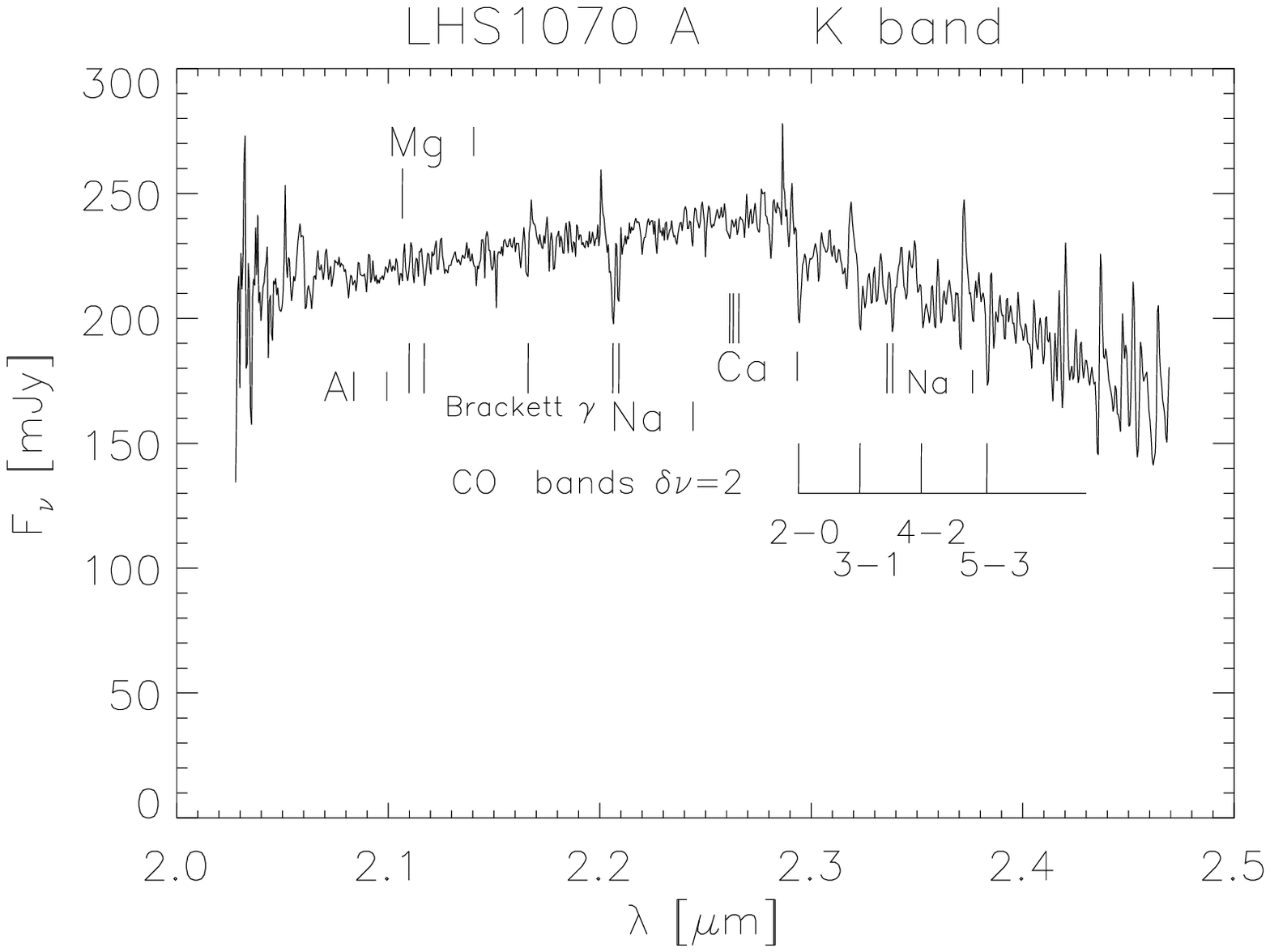}
   \includegraphics[width=6cm]{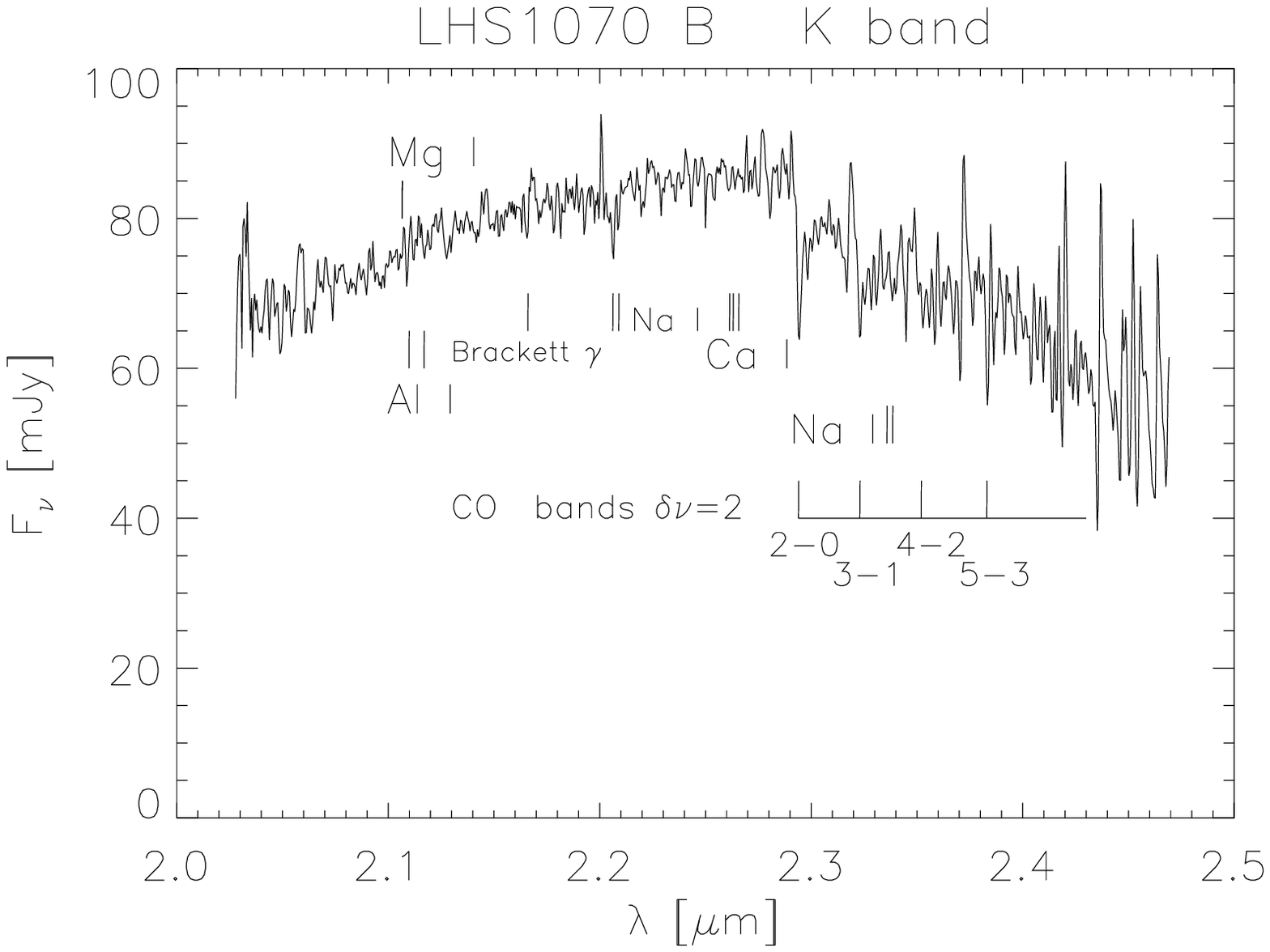}
   \includegraphics[width=6cm]{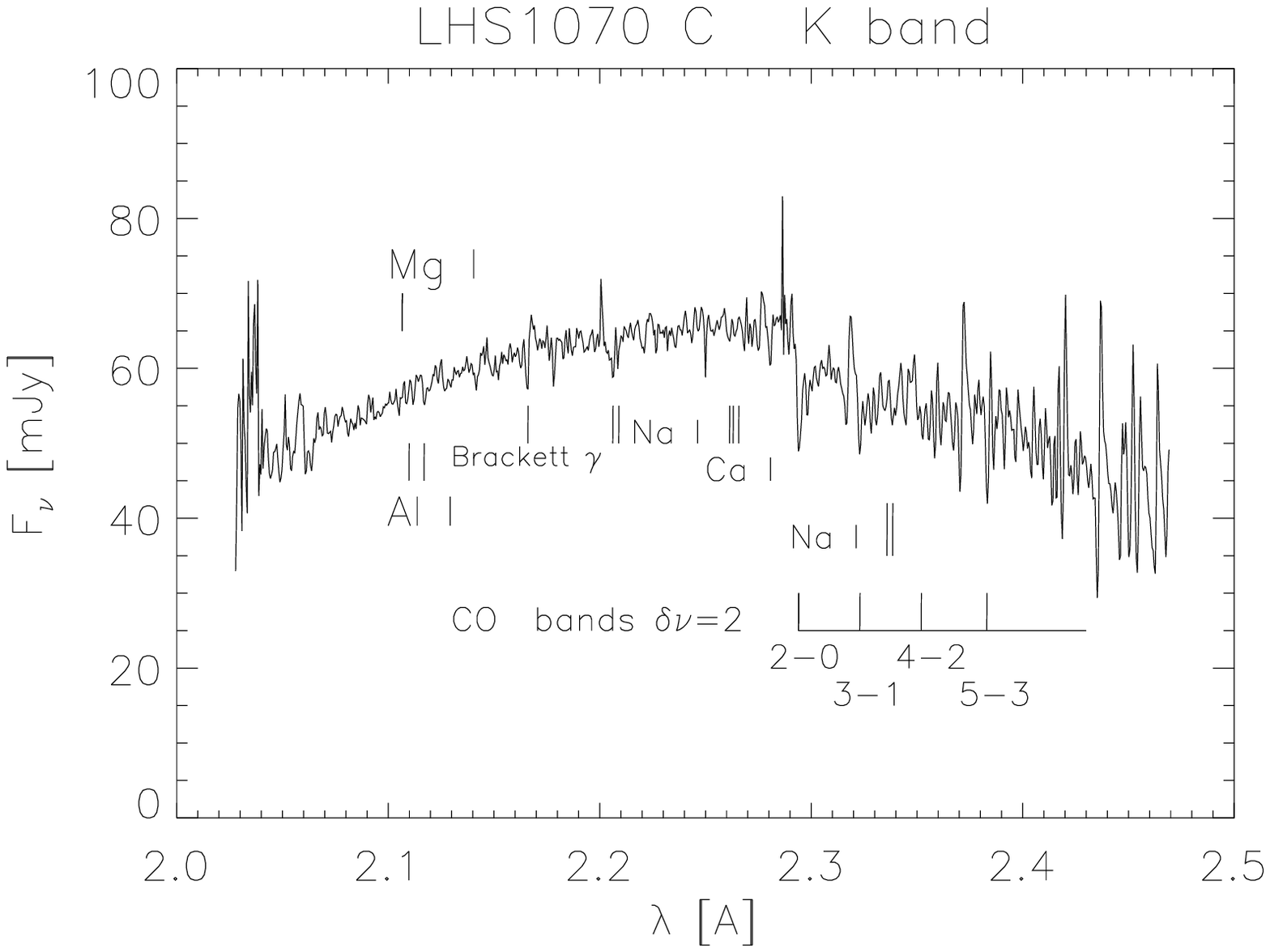}
   \caption{VLT (NACO) spectra of all the components in $J$, $H$ and
     $K$ bands with atomic and molecular lines indicated.} 
         \label{Fig3}
   \end{figure*}

Figure~\ref{Fig3} shows the NACO spectra obtained at the VLT of the
three individual components in the $J$ (upper panel), $H$ (middle
panel), and $K$ (lower panel) band for components A, B, and C (from
left to right). 
The main expected features are also indicated. The Paschen Beta and CaI lines can be seen in the $J$-band spectra
with ``equivalent widths'' of 2-2.5\,$\AA$ and 1.5-2\,$\AA$, respectively. 
The $H$-band spectra contain many relatively weak absorption features,
which defy definite identifications, with possible exception of Mg (1.711 $\mu$m), OH (1.689 $\mu$m), and Al (1.675 $\mu$m). 
H$_2$O bands define the shape of the $J$ and $H$ band peaks. Water
absorption is most obvious in the $J$-band at 1.33\,$\mu$m and
strengthens through the later types: the flux ratio between the peak
and the minimum of the absorption band increases from 1.09$\pm$0.01
for component A to 1.30$\pm$0.01 and 1.39$\pm$0.01 for the cooler components.
The $K$-band spectra of the three components show strong CO bands and 
more or less pronounced atomic features.
The NaI lines weaken from the hotter component A to the cooler
components B and C as dust forms. 

\section {Model atmospheres}
\label{mod}

The atmospheres of low-mass stars are mostly composed of molecular
hydrogen and CO, and their spectra characterized by strong TiO bands
in the optical, water vapor bands in the infrared, chromospheric
activity, flares, magnetic spots, and planet like properties
\citep{Allard1997}. The discovery of the formation of dust clouds in M
and L dwarfs makes it even more challenging to understand their
spectral properties. An effective temperature of 2600\,K and below is
sufficiently low to give rise to enough silicate dust formation in the
photospheric layers to affect the spectral properties of late type M
dwarfs. These grains produce a ``veiling'' in the optical by dust
scattering and an important greenhouse effect (redistribution of flux
to the infrared) which strongly influence the infrared spectral
properties. 
\cite{Tsuji1999,Allard2001} treated the dust formation in the low-mass
stars in pure Chemical Equilibrium (hereafter CE). CE yields the
formation of  condensates from zirconium dioxide and silicates
(Mg$_2$SiO$_4$) to refractory ceramics (CaTiO$_3$, Al$_2$O$_3$), salts
(CsCl, RbCl, NaCl), and ices (H$_2$O, NH$_3$, NH$_4$SH$_4$) depending
on the temperature of the atmosphere from M through T and Y spectral
types \citep{Allard2001,Lodders2006}. 
\cite{Helling2008c,Helling2008b,Allard2012a} explored the properties
of the formation of the dust clouds in low-mass stars by determining
the radial distribution and average size of the grains.  

We compared the flux- calibrated spectroscopic data of LHS\,1070 with
synthetic spectra computed from three recent atmosphere models for
cool stars. These models are described below. In order to match the
observed spectra, the synthetic spectra have been scaled by the
dilution factor (R/d)$^{2}$ where $d$ is the distance of the system
from \cite{Costa2005} and the stellar radius $R$ is a free parameter
ranging from 0.096 $R_\odot$ to 0.142 $R_\odot$ at a step of
0.002. The radius resulting from the fits  can then be compared to
predictions from stellar evolutionary models. 

\subsection{BT-Dusty and BT-Settl}

Recently, the preliminary results of the BT-Cond/Dusty and BT-Settl
models based on the \cite{Asplund2009} solar abundances have been
published in a review by \cite{Allard2012a}, and those grids
distributed via the \texttt{PHOENIX} web simulator
(http://phoenix.ens-lyon.fr/Grids/BT-Settl/AGSS2009/).  For this
paper, we have developed a revision to these BT-Dusty and BT-Settl
models based on the \cite{Caffau2011} solar abundances using slightly
revised atomic and molecular opacities as well as cloud physics. Their
detailed publication is in preparation. We summarize below the
differences between the current version of the models and the
published versions of \cite{Allard2001,Allard2012a}. 

The BT appellation stands for the Barber and Tennyson so-called BT2
water vapor line list \citep{BT2H2O}. The models are provided in
several versions addressing different limiting treatment of the cloud
physics. The Dusty and Cond appellation, as defined by
\cite{Allard2001}, refer to a CE treatment of dust formation where
dust opacities (assuming spherical grains with interstellar grain size
distribution) are ignored in the Cond models.  The BT-Dusty and
BT-Cond models reproduce the color properties of the 2001 AMES Dusty
and Cond models in the brown dwarf regime, with minor differences
relating to revised opacities for 
H$_2$O, TiO \citep{Plez1998}, VO and most of the hydrides, detailed
damping profiles for the alkali lines \citep{Allard2007}, the damping
constants of molecular lines \citep{Homeier2005}, new line lists for CH$_4$
\citep[STDS,][]{Homeier2005}, CO$_2$ \citep[CDSD,][]{Tashkun2004}, and CIA
\citep{Borysow2001,Abel2011} to mention the most important changes). 
We reserve the NextGen appellation, as chosen for the
\cite{Hauschildt1999} models, for corresponding pure gas phase
models. 

The BT-Settl models described by \cite{Allard2012a}, on the other hand,
include gravitational settling which is ignored in Dusty models, and
which involves a cloud model. The growth of the grains is governed by
the supersaturation ratio of the gas and is triggered by the actual
collision rates between grains and gas molecules and, therefore,
depends on the grain surface.  A breakthrough was therefore achieved,
compared to earlier versions
\cite[]{Allard2003,Allard2007,Reiners2007b,Helling2008a}, by
calculating locally the supersaturation instead of assuming the fixed
conservative value proposed by \cite{Rossow78}. The cloud model used
in the BT-Settl models is based on \cite{Rossow78} and accounts for
nucleation, condensation, supersaturation, gravitational settling or
sedimentation, and advective mixing. The latter includes convective
mixing based on the mixing length theory in the convection zone, an
exponential overshoot according to \cite{Ludwig2006}, and the gravity
waves according to \cite{Freytag2010}. The latter dominate the mixing
in the uppermost atmosphere layers, i.\,e.\ the regions where clouds
first begin to form in late M dwarfs. The grains are still considered
spherical but the grain sizes (a unique or mean value per atmospheric
layer) and grain number densities are determined by the comparison of
the different timescales, and thus vary with depth to reach grain
sizes of a few times the interstellar values at the cloud base.  The
BT-Settl models do not enforce grains to be in equilibrium with the
gas phase as is the case of the Dusty and Cond models. However, the
gas phase opacities reflects the depletion of elements from the gas
phase caused by grain growth. The numerical solving on-the-fly of the
gas phase CE in the BT-Settl models (unlike what is often done by
other authors) costs computing time but allows to account for the
cooling history of the atmospheric layers. 

For the current BT-Settl model version, we have additionally
explored the effects of nucleation in our cloud model by assuming a
constant nucleation rate of 1\,cm$^{-3}$. This acts as a limiting
factor to the creation, sedimentation and depletion of grains and
thus leads to the persistence of grains higher up in the atmospheric
structure than assuming unlimited grain production. 
We are using in general atomic damping constants according to
\citet{unsoeld55} with a correction factor to the van der Waals widths
of 2.5 \citep{Valenti1996}, van der Waals broadening of molecular
lines with generic widths according to \citet{Homeier2003}. 
More accurate broadening data for neutral hydrogen collisions by
\citet{Barklem2000} have been included for several important atomic
transitions such as the Alkali, the Ca\,\textsc{i} and Ca\,\textsc{ii}
resonance lines. 

Grains are assumed here to be spherical and non-porous, and their
Rayleigh to Mie scattering and absorptive properties are considered
for 55 types of material including the species mentioned above and
presented in \cite{Allard2001} and \cite{Ferguson2005}, plus MnSiO$_3$
and Na$_2$O$_5$Si$_2$. 
The grain opacities are computed in each layer for pure spherical
grains using the radius determined by the cloud model (assuming it
represents an average radius of the grains in each layer) and the
material density as measured in the laboratory. The opacity
contributions of the various grain species present in each layer are
finally summed.  

The model atmospheres and synthetic spectra are computed with the
\texttt{PHOENIX} radiative transfer code \citep{Allard1990,Allard1995,Allard2001} 
using hydrostatic equilibrium, convection based on the mixing length
theory and a mixing length which varies from 2.2 to 1.6 from brown
dwarfs to the Sun according to results of radiation hydrodynamical
simulations \citep{Ludwig1999,Ludwig2002,Ludwig2006}, spherically symmetric
radiative transfer using radii provided by published evolution models,
micro-turbulence velocities from radiation hydrodynamical simulations
\citep{Freytag2010}, and the latest solar abundances by
\cite{Caffau2011}. 
 
The synthetic spectra are provided over the entire spectral range of
interest at a spectral resolution of 0.05\,\AA in the optical and
0.1\,\AA in the near-IR. 
For this paper, we explored these model grids with parameters described as follows:
\begin{itemize} 
\item $T_\mathrm{eff}$ from 2000\,K to 3100\,K with 100\,K step, as expected for mid-M to L dwarfs,
\item $\mathrm{log}\,g  = 4.5, 5.0$, and 5.5 dex, 
\item $[M/H] = -1.5, -1.0, -0.5, 0.0, +0.3$, and $+$0.5 dex,
\end{itemize}

\subsection{MARCS}

The MARCS  code \citep{Gustafsson2008} assumes hydrostatic equilibrium, Local
Thermodynamic Equilibrium (LTE), chemical equilibrium, homogeneous plane-parallel stratification, and the conservation of the total flux (radiative plus convective; the convective flux being computed using the local mixing length recipe). The  radiation field used in the model generation is
calculated by assuming absorption from atoms and molecules by opacity sampling at approximately 100\,000 wavelength points over the wavelength range $1300\,\mbox{\AA} $--$ 20\,\mbox{$\mu$m}$.  

The code used for calculating the synthetic spectra is BSYN v. 7.12 which is based on routines from the MARCS code.  The atomic line list used
in our calculations is compiled from the VALD I database \citep{Kupka1999} updated according to \cite{Gustafsson2008}. The molecular line lists
include CO, SiO, TiO, ZrO, VO,  OH, H$_2$O,  CN, C$_2$, NH, CH, AlH, SiH, CaH \citep[see references in][]{Gustafsson2008}, MgH \citep{Skory2003}, FeH \citep{Dulick2003}, and CrH \citep{Burrows2002}. Up-to-date dissociation energies and
partition functions are used. The basic chemical composition adopted is that of the Sun as listed by \cite{Grevesse2007}. The synthetic spectra were calculated in plane parallel symmetry. A constant micro-turbulence velocity of
$2\,\mathrm{km\,s}^{-1}$ is assumed. 
The most important differences to the BT models are the different opacities sources, the solar abundances, and the fact that MARCS is a pure gas phase model with no dust formation. 

Synthetic spectra are calculated for the wavelength region of $0.50-2.53\mu$m, with a resolution of  $R=600 000$. 
We used a grid of MARCS model atmosphere which spans the parameters as follows: 
 
\begin{itemize} 
\item $T_\mathrm{eff}$ from 2500\,K to 3100\,K with 100\,K step (lower temperatures are not available),
\item $\mathrm{log}\,g = 4.5, 5.0$, and 5.5 dex,
\item $[M/H]$ from $-$0.5 to 0.25 dex with 0.25 dex step.
\end{itemize}

\subsection{DRIFT}

The \texttt{DRIFT-PHOENIX}  model atmosphere code has been developed by \cite{Dehn2007, Helling2008b, Witte2009}. 
Both the DRIFT code by \cite{Helling2008c} and the BT-Settl code
return a consistent dust cloud structure with corresponding opacities
and the altitude-dependent depletion and redistribution of gas phase
abundances, which feed back on both the thermodynamical structures and
the radiation field. An iteration of these methods allows the
determination of stationary atmosphere and dust cloud properties and
yields the respective synthetic spectra. But solely in the case of the
DRIFT code, the dust formation takes place via the formation of seed
particles and their subsequent growth or evaporation, solving from top
to bottom of the atmosphere. The BT-Settl models in contrast solve the
timescale equations to calculate the depletion of refractory elements
from bottom to top. 

As in the BT-Settl model, the DRIFT model assumes dirty grains mixed
according to the composition of each atmospheric layer. While the
BT-Settl model assumes dirty grains in the timescales equations to
calculate the growth and settling of the grains, it only sums the
opacity contributions of each species in each layer as for an ensemble
of pure grains.
The DRIFT model instead uses composite optical constants calculated in
effective mean theory from the composition of the grains
\citep{Bosch2000}. The resulting absorption and scattering 
properties of the grains are therefore different than those of the
BT-Settl models, possibly producing more opaque clouds. However, since
the opacities are dominated by atomic and 
molecular opacities over most of the spectral distribution in this
spectral type range, the impact of those differences are difficult to
identify. The largest differences between the BT-Dusty, BT-Settl and
DRIFT models are the differences in the local number density, the size
of dust grains, as well as their mean composition, which are the
direct results of the cloud model approach. 

The DRIFT model also considers only seven of the most important solids
(TiO$_2$, Al$_2$O$_3$, Fe, SiO$_2$, MgO, MgSiO$_3$, Mg$_2$SiO$_4$)
made of six different elements, found to form below an effective
temperature of 2800\,K.  Note that the CE, which determines the 
composition of the solid species included in the BT-Settl model, does
not give the same list. Among the most important species, forming in
the BT-Settl model below $T_\mathrm{eff}=2900$\,K, are additionally
ZrO$_2$, CaTiO$_3$, CaSiO$_3$, Ca$_2$SiO$_4$, Ca$_2$Al$_2$SiO$_7$,
Ca$_2$MgSi$_2$O$_7$, CaMgSi$_2$O$_6$, Ti$_2$O$_3$, Ti$_4$O$_7$, 
Fe, Ni, VO, V$_2$O$_3$, MgTiO$_3$,
MgTi$_2$O$_5$, MgAl$_2$O$_4$, Al$_2$Si$_2$O$_{13}$. 
The DRIFT model includes, similarly to the BT-Settl model, mixing by
convection and overshooting by assuming an exponential decrease in
mass exchange frequency in the radiative zone \citep{Woitke2004}. However
it neglects the contribution of the gravity waves included in the
BT-Settl model.  The model code \texttt{DRIFT-PHOENIX}  has been
discussed in more detail by \cite{Witte2009}. 

We used a grid of DRIFT model atmosphere which spans the parameters as follows:
 
\begin{itemize} 
\item $T_\mathrm{eff}$ from 2200\,K to 3100\,K with 100\,K step,
\item $\mathrm{log}\,g= 4.5, 5.0$, and 5.5 dex, 
\item $[M/H]$ from $-0.5$ to 0.5 with 0.5 dex step.
\end{itemize}

\section {Physical parameters determination and results}
\label{result}

The first analysis of LHS\,1070 by spectral synthesis was made by \cite{Leinert1998} using the FOS spectra and the pure gas phase NextGen model atmospheres \citep{Allard1997,Hauschildt1999}. But the quality of the fits, even for the A component was disappointing, especially below $0.72~{\mu}$m, and the $T_\mathrm{eff}$ of the B and C components were strongly overestimated (2700K) mainly due to the absence of dust formation. \cite{leinert2000} have therefore used the AMES-Dusty models \cite[]{Allard2001} to re-analyse the LHS\,1070 system. However if the quality of the fits and precision of the $T_\mathrm{eff}$ for the B and C components were clearly improved, this was still clearly not the case of the M dwarf primary.  The stellar parameters obtained with the assumption of a distance of 8.8 pc are summarized:
\begin{itemize}
\item Component A: $T_\mathrm{eff}=2950$\,K, $\mathrm{log}\,g=5.3\pm0.2$, $[M/H]=0.0$
\item Component B: $T_\mathrm{eff}=2400$\,K, $\mathrm{log}\,g=5.5\pm0.5$, $[M/H]=0.0$
\item Component C: $T_\mathrm{eff}=2300$\,K, $\mathrm{log}\,g=5.5\pm0.5$, $[M/H]=0.0$
\end{itemize}

In the following, we derive the stellar parameters 
using more recent atmosphere models and spectroscopic informations covering both the
optical and IR ranges. 
Metallicity and gravity are
determined from peculiar spectral features, whereas effective temperature and
radius are constrained from the overall shape of the spectra, following the different steps: (i) a first $\chi^2$ minimization is performed 
on the overall spectra considering effective temperature, radius, metallicity, and gravity as free parameters. It gives a first guess for the 
parameter space of each component, (ii) we looked for peculiar spectral features that are mainly sensitive to metallicity (see Sect.~\ref{feh})
or gravity (see Sect.~\ref{logg}) to refine these two parameters, (iii) we fixed these parameters to perform
another $\chi^2$ minimization and derive effective temperature and radius  (see Sect.~\ref{teff}). At each step we checked that the resulting value  
is not sensitive to changes on the value of the other parameters.

Age is
estimated from kinematics and rotation. Before entering the details of our study,
we also summarize our results in Table~\ref{tab:3}. Note that these values have
been obtained by assuming a distance of 7.72 pc \citep{Costa2005}
whereas \cite{leinert2000} used the higher value of 8.8 pc. 

\begin{table*}
\caption{Derived parameters for the LHS\,1070 system. The luminosity $L$ is computed from the radius and the effective temperature.}
\label{tab:3}
\centering
\begin{tabular}{c c c c c}
\hline
\multicolumn{5}{c}{Component A} \\
\hline
Model & $T_\mathrm{eff}$ & $\mathrm{log}\,g$  &Radius  & log($L$) \\
 &  (K) & (cm\,s$^{-2}$) &($R_\odot$ ) &  ($L_\odot$)\\
\hline
BT-Dusty & 2900$\pm$100 & 5.0  & 0.134$\pm$0.005& $-$2.93$\pm$0.090\\
BT-Settl & 2900$\pm$100 & 5.0  & 0.134$\pm$0.005& $-$2.93$\pm$0.090\\
MARCS   & 2900$\pm$100& 5.0 & 0.136$\pm$0.005& $-$2.92$\pm$0.090\\
DRIFT   &2900$\pm$100 &5.0& 0.136$\pm$0.005& $-$2.92$\pm$0.031\\
isochrones & 2953 & 5.2 & 0.140 & $-$2.88 \\ 
\hline
\multicolumn{5}{c}{Component B} \\
\hline
Model & $T_\mathrm{eff}$ & $\mathrm{log}\,g$  &Radius  & log($L$) \\
 &  (K) & (cm\,s$^{-2}$) &($R_\odot$ ) &  ($L_\odot$)\\
\hline
BT-Dusty & 2500$\pm$100 & 5.0  & 0.102$\pm$0.004& $-$3.43$\pm$0.105\\
BT-Settl & 2500$\pm$100 & 5.0  & 0.102$\pm$0.004& $-$3.43$\pm$0.105\\
MARCS   & 2600$\pm$100& 5.0 & 0.098$\pm$0.002  & $-$3.39$\pm$0.086\\
DRIFT   &2400$\pm$100 &5.0&0.106$\pm$0.005& $-$3.46$\pm$0.044\\
isochrones & 2432 & 5.3 & 0.104 & $-$3.48\\
\hline
\multicolumn{5}{c}{Component C} \\
\hline
Model & $T_\mathrm{eff}$ & $\mathrm{log}\,g$  &Radius  & log($L$) \\
 &  (K) & (cm\,s$^{-2}$) &($R_\odot$ ) &  ($L_\odot$)\\
\hline
BT-Dusty & 2400$\pm$100 & 5.0  & 0.098$\pm$0.002& $-$3.53$\pm$0.090\\
BT-Settl & 2400$\pm$100 & 5.0  & 0.098$\pm$0.002& $-$3.53$\pm$0.090\\
MARCS   & 2500$^a$$\pm$100& 5.0 & 0.100$\pm$0.002& $-$3.44$\pm$0.090\\
DRIFT  &2300$\pm$100 &5.0&0.102$\pm$0.005& $-$3.57$\pm$0.034\\
isochrones & 2234 & 5.3  & 0.098 & $-$3.68\\
\hline
\end{tabular}

$^a$ Note that this value might be meaningless as there is no cooler available model.
\label{tab:3}
\end{table*}

\subsection{Spectral type}
A well-defined spectral classification for the M dwarfs helps in the calibration of the temperature of the late type stars and for the definition of
the end of the main sequence on the HR diagram. We have derived spectral indices and spectral types for all three components. For the early
M-dwarf, i.e. the primary, we used the classification scheme based on the TiO and CaH band-strengths, as defined by \cite{Reid1995}. For the late M-dwarfs, we also used the PC$_3$ index defined by \cite{Lepine2003a,Hawley2002,Martin1999}. We have computed the spectral index H$_2$O-K in the near IR K-band defined by \cite{Covey2010}  and used the spectral-type vs index relation from \cite{Rojas-Ayala2010}.
The spectral indices and corresponding spectral types are given in Table
\ref{tab:4}. The obtained spectral classification for components B and C is one subclass higher as compared to \cite{leinert2000}.
The spectral types obtained in the K-band differs from the optical indices by up to two subclasses showing inconsistency on the spectral type versus spectral index relations defined from the optical and near-IR spectra. Here we adopt the spectral type obtained from optical spectral indices.

\begin{table}[ht]
\caption{Spectral index values and derived spectral type computed from
  TiO and CaH band-strengths for component A and from the PC3 index
  for components B and C.} 
\label{tab:4}
\centering
\begin{tabular}{c c c c c}
\hline
Band&Indices&Spectral type\\
\hline
\multicolumn{3}{c}{Component A} \\
\hline
TiO$_5$& 0.211& M5.5\\
CaH$_2$& 0.281& M5.5\\
CaH$_3$& 0.557& M6\\
H$_2$O-K&0.829&M6.5\\
Adopted value & &M5.5\\
\hline
\multicolumn{3}{c}{Component B} \\
\hline
PC$_3$    & 2.305 &M9.5 \\
H$_2$O-K&0.791&M8.5\\
Adopted value & &M9.5 \\
\hline
\multicolumn{3}{c}{Component C} \\
\hline
PC$_3$    &  2.608 &L0 \\
H$_2$O-K&0.798&M8\\
Adopted value & &L0 \\
\hline
\end{tabular}
\label{tab:4}
\end{table}

\subsection{Metallicity}
\label{feh}

\begin{figure}[ht]
   \centering
   \includegraphics[width=8.6cm,clip=]{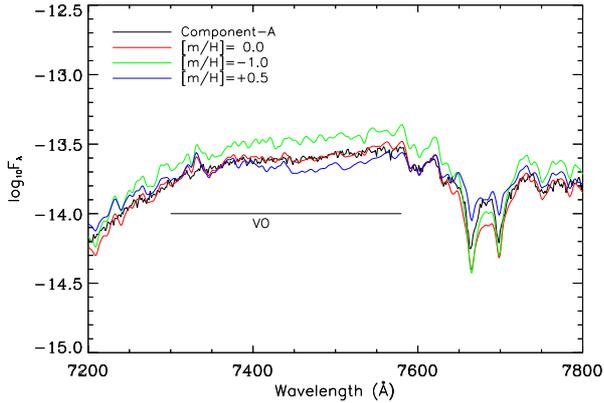}
      \caption{VO band observed in the primary (black) compared to the BT-Settl model at 2900 K, $\mathrm{log}\,g= 5.0$, $R = 0.134 R_\odot$ 
                for different metallicities.}
         \label{Fig7}
   \end{figure}

In order to estimate the metallicity of the system, we looked for special features in the spectra that are mainly sensitive to the metallicity.
The main indicator of metallicity is the VO absorption band at 7300~{$\AA$}~-~7600~$\AA$. It is well reproduced at solar metallicity for all components, and shown in Fig.~\ref{Fig7}  for the primary with the BT-Settl model. 
We checked that the same metallicity is found when changing the other parameters.
This solar metallicity can also be inferred from  the NaI, CaI, and H$_2$O features in the K-band of the primary using the calibration determined by \cite{Rojas-Ayala2010}.
In the following, we adopt this $[M/H]=0$~dex value.

\subsection{Gravity}
\label{logg}

\begin{figure}[ht]
   \centering
   \includegraphics[width=8.6cm]{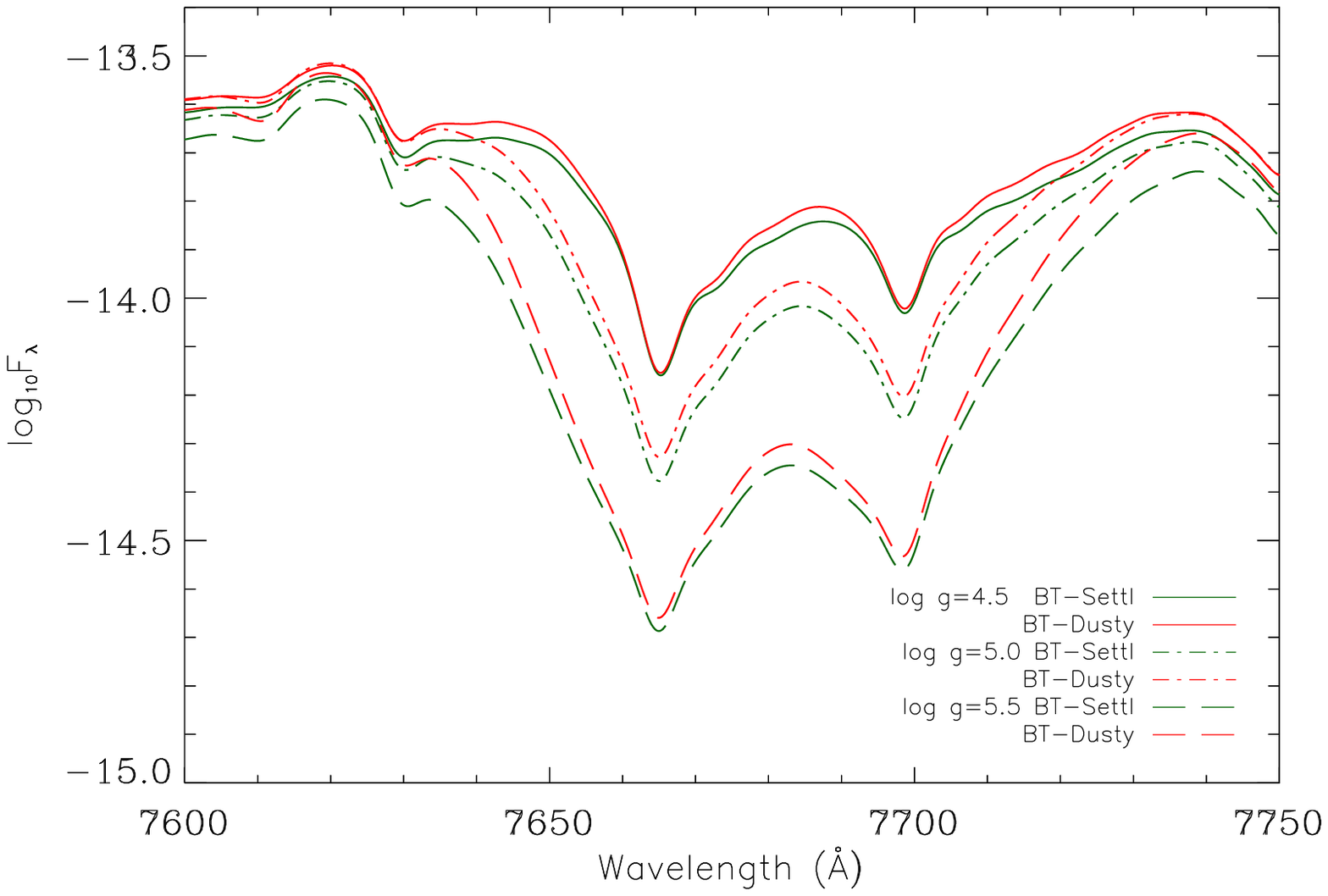}
   \includegraphics[width=8.6cm]{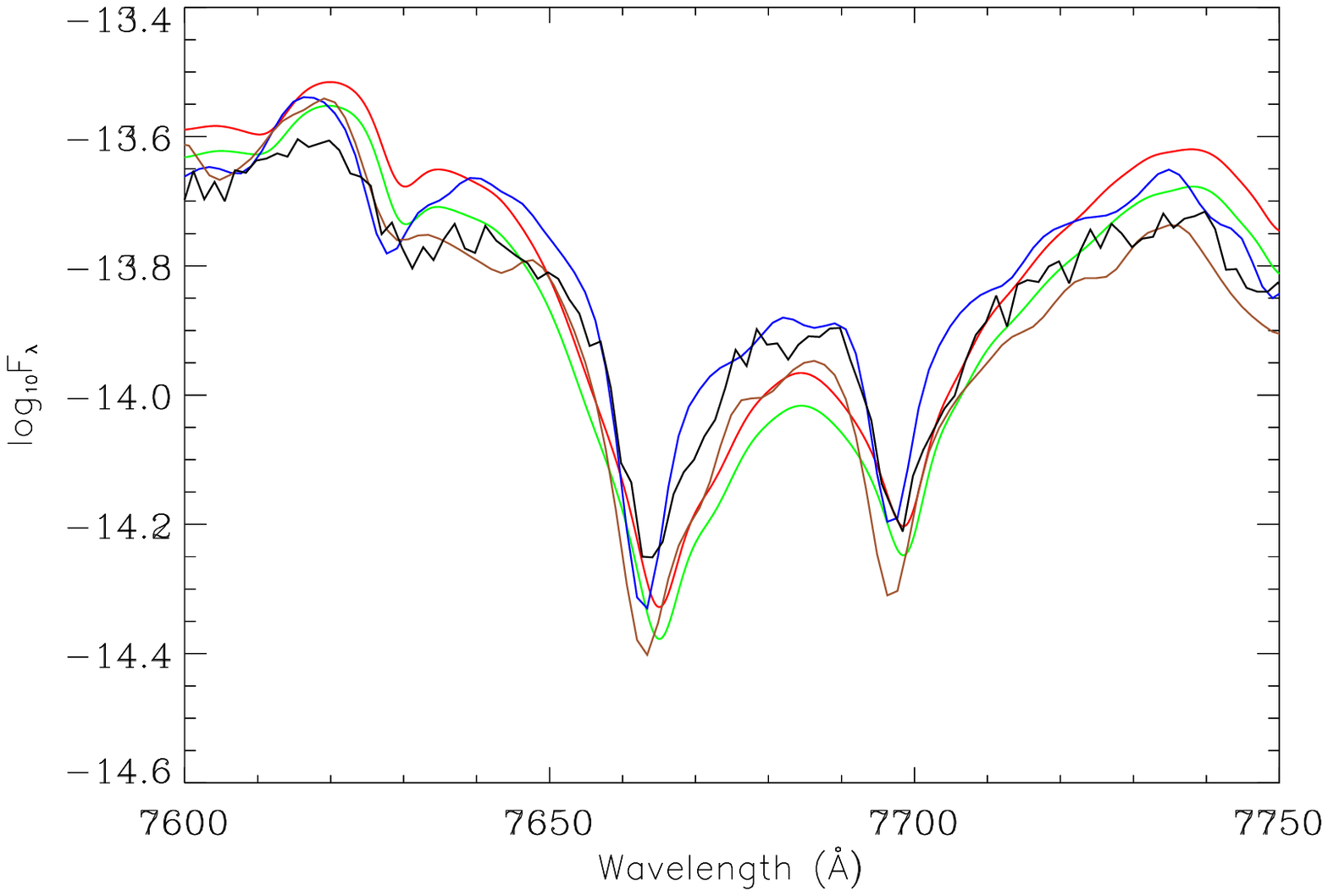}
      \caption{Right: BT-Settl and BT-Dusty models for an effective
        temperature of  2900\,K and varying log\,$g$. The effect of
        gravity and the pressure broadening of the K\,I doublet is
        clearly visible. The details of the dust treatment only cause
        negligible differences at this $T_\mathrm{eff}$. 
Left: K\,I doublet as observed for the primary (black) compared to the 
BT-Dusty (red), BT-Settl (green), DRIFT (blue), and MARCS
(brown) models with $\mathrm{log}\,g = 5.0$. 
}
\label{Fig5}
   \end{figure}

The surface gravity can be estimated by analyzing the width of atomic
lines such as the K\,I and Na\,I~D doublets, as well as the relative
strength of metal hydrides bands such as those of CaH. The K\,I
doublet at 7665\,$\AA$ and 7699\,$\AA$ is a particular useful gravity
discriminant for M stars. 
Figure~\ref{Fig5} (left panel) shows the gravity effects on the
strength and pressure broadening of the K\,I lines as modeled by the
\texttt{PHOENIX} BT-atmospheres. The overall line strength (central
depth and equivalent width) increases with gravity as the
decreasing ionisation ratio due to the higher electron pressure leaves
more neutral potassium in the deeper atmosphere. The width of the
damping wings in addition increases due to the stronger pressure
broadening, mainly by H$_2$, He and H\,I collisions. 

\begin{figure}[ht]
   \centering
   \includegraphics[width=8.6cm]{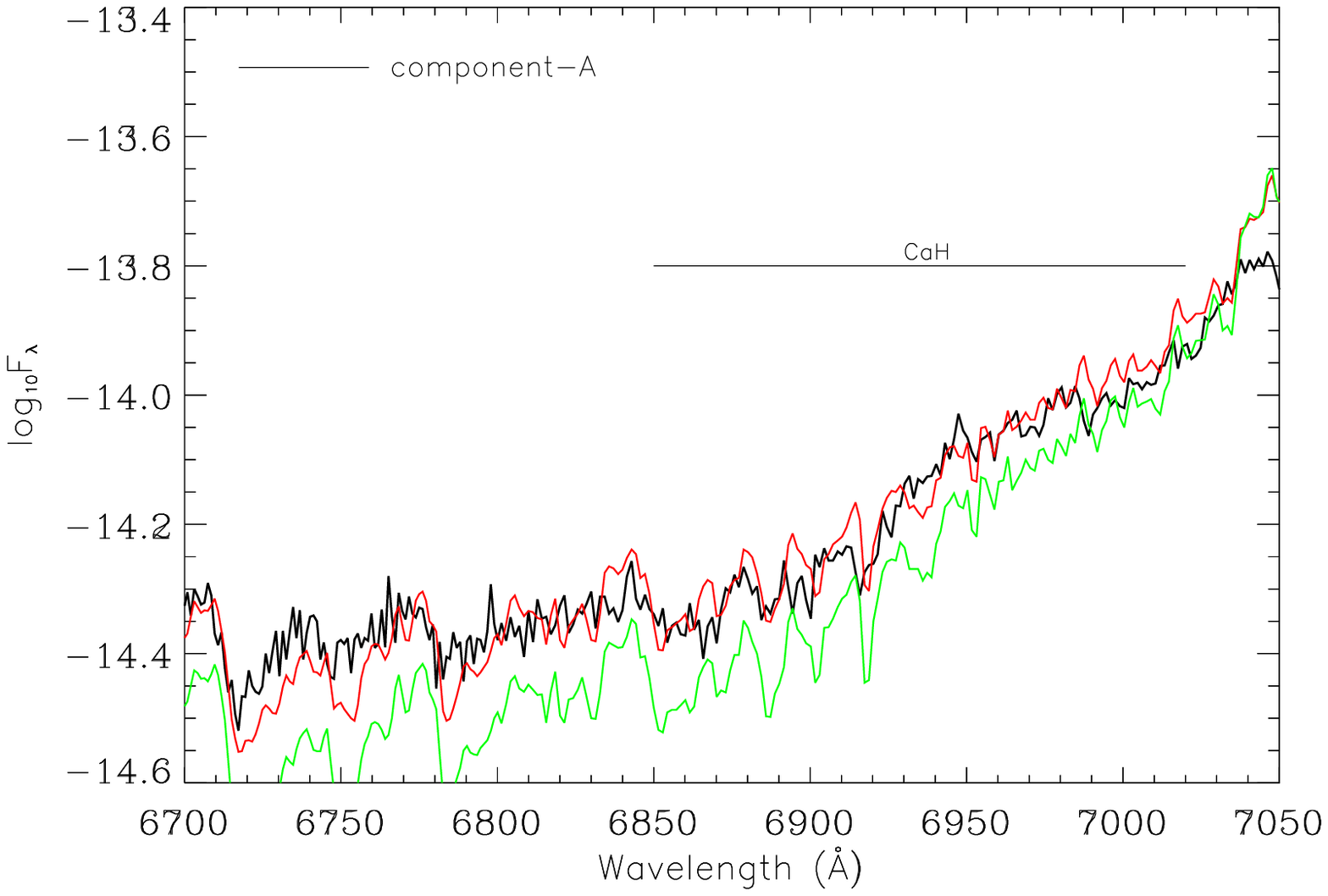}
      \caption{CaH molecular bands in the spectrum of the primary (black) compared to the BT-Settl model at 2900 K, $\mathrm{log}\,g= 5.0$ (red) and $\mathrm{log}\,g=5.5$ (green).}
         \label{Fig6}
   \end{figure}

Figure~\ref{Fig5} (right panel) shows the determination of gravity for
the component A from the K\,I doublet using the BT-Settl, MARCS, and
DRIFT models. 
The determined log\,g values are given in Table~\ref{tab:3}. The best agreement is obtained with $\mathrm{log}\,g = 5.0$~dex for
all the atmospheric models.
This value is confirmed by the CaH
molecular bands (Figure~\ref{Fig6}). 
We checked that the same metallicity in found when changing the other parameters
and adopt the value $\mathrm{log}\,g = 5.0$~dex in the following.
For comparison, the gravities inferred
from the masses found from the orbits of the system
\citep{Seifahrt2008} are also listed, which are also in good agreement
with the values found by fitting.

\subsection{Effective temperature and radius}
\label{teff}

We performed a second $\chi^2$ minimization by adopting the metallicity and gravity derived in Sect.~\ref{feh} and \ref{logg} and refined the effective temperature and 
the radius by comparing the overall shape of the observed and synthetic spectra.
As opposed to the studies mentioned previously in which the best
fit was found by trial and error, in this paper we derive the
effective temperature and radius by performing a $\chi^2$ minimization
technique.  
For this purpose, our approach was to first convolve the synthetic spectrum with a Gaussian 
kernel at the observed resolution and then rebin the outcome with the observation. 
For each of the observed spectra we have calculated the reduced
$\chi^2$ value by comparing these spectra, taking into account their uncertainties estimated from the reduction procedures (see Sect.~\ref{obs-spec}), with the grids of synthetic
spectra in the wavelength range between 4500~\AA\ to 2.4~$\mu$m. We
have excluded the spectral region below 4500~{\AA} due to the low $S/N$
ratio of the observed spectra. The number of data points used for the $\chi^2$ computation is thus 1487 in the optical and 204 in the infrared.

\begin{figure*}[ht]
   \centering
   \includegraphics[width=8.6cm]{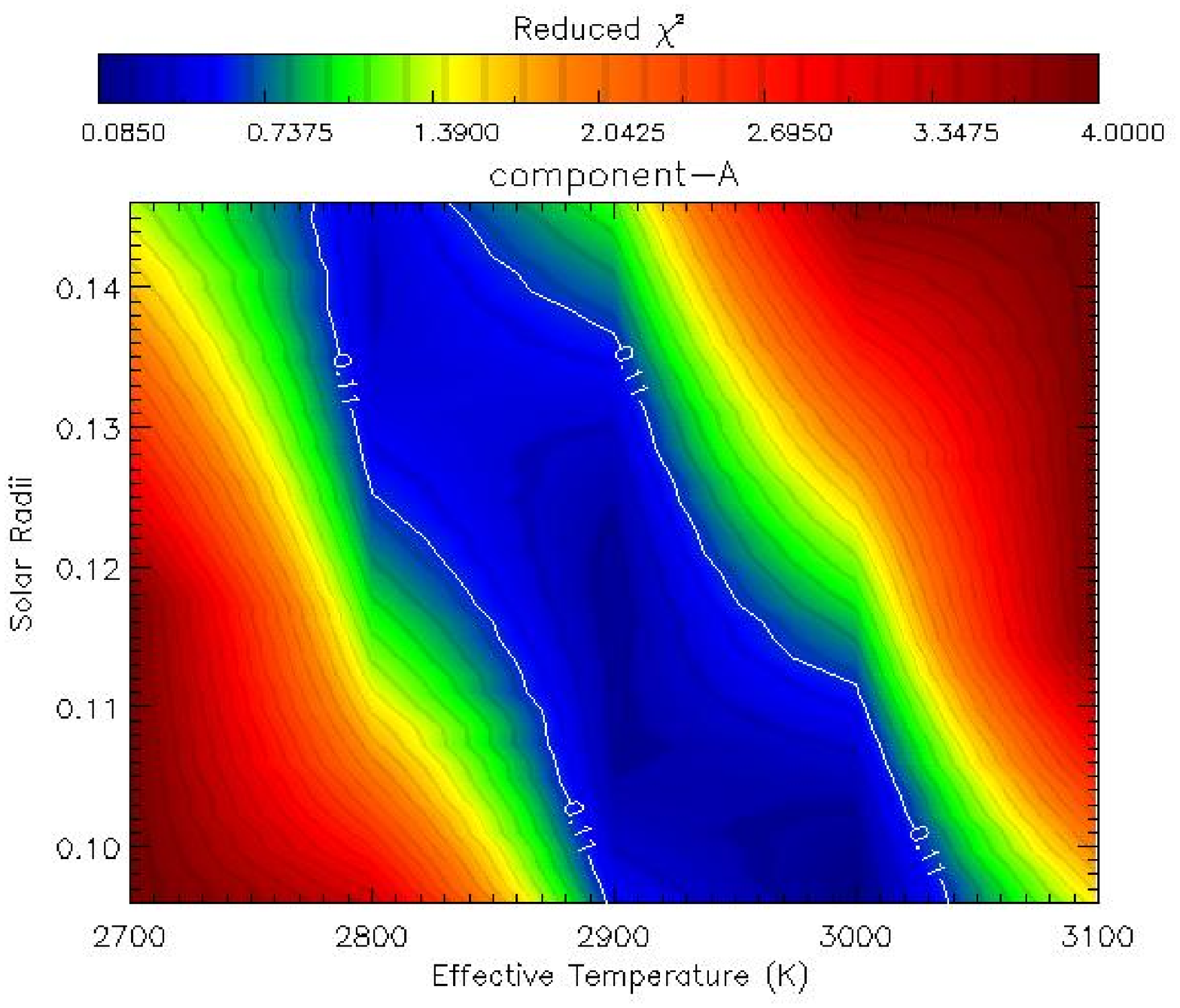}
   \includegraphics[width=8.6cm]{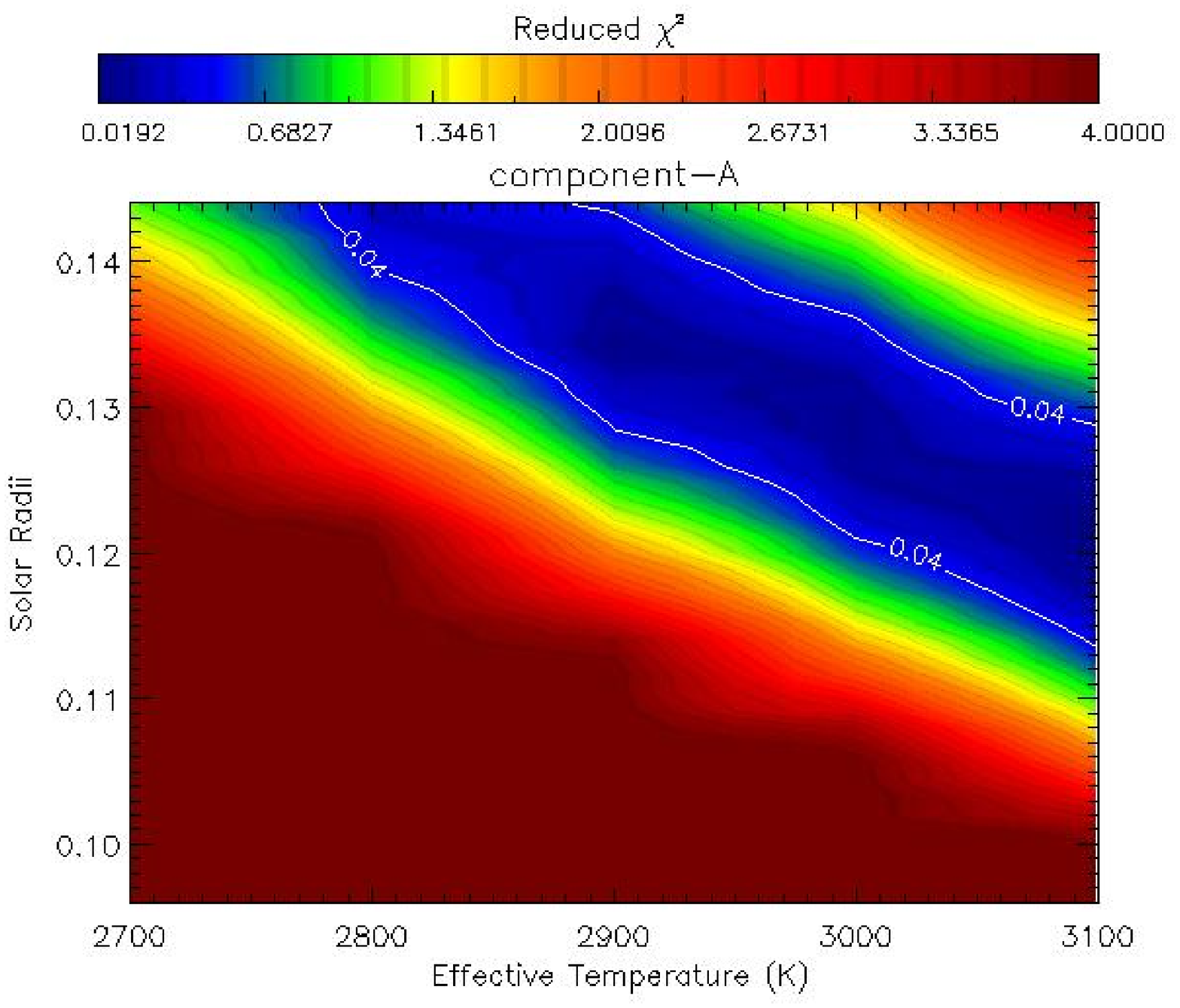}
      \caption{ Left: $\chi^2$ map computed for component A using the optical spectrum and the BT-Dusty model. Right: same using the IR spectrum. All values inside the white contour give an acceptable fit (checked by eye).  The adopted common solution to both $\chi^2$ maps is: $T_\mathrm{eff}=2900$ K and a radius of 0.134 $R_\odot$.}
         \label{Fig:4}
   \end{figure*}

In a second step, a reduced $\chi^2$ map has been obtained for each
component in the optical and in the infrared as a function of
temperature and radius. Such a map is shown in Fig.~\ref{Fig:4} for
the primary using the BT-Dusty model. The $\chi^2$ minimum value is given on the left part of the colour bar.  The parameter space which gives an acceptable solution around the minimum $\chi^2$ 
valley is within the white contour, defined by visual inspection. The  $\chi^2$ value at this significance level is indicated along the white contour.
Error bars are derived from this contour.

We have identified on the contour maps the possible solutions in the
optical and IR.  The adopted values of effective temperature and radius 
are the common intersection between the solutions found in 
the optical and IR.

The solutions were finally inspected by comparing it with the
observed spectra. The same procedure has been used with BT-Settl,
MARCS and DRIFT model grids. Differences of 100\,K to 200\,K in the
$T_\mathrm{eff}$ determination (see Table~\ref{tab:3}) are found for
the B and C components depending on the model used, whereas all models
agree on the effective temperature of the primary. 

\subsection{Age and mass}
With its high velocity component of $W=43$\,km/s perpendicular to the
galactic plane based on \cite{Basri1995}, LHS1070 has been considered
as part of the old disk population with an age of several
Gyrs. \cite{Reiners2007a} refined this estimate to about 1~Gyr based
on measured rotation velocities of its components using a modified
Skumanich braking law \citep{Skumanich1972}.  They do not exclude
however that the braking law may also have to be changed in its absolute
time scale, which could increase this estimate.  

By an orbital fit, \cite{Leinert2001} have computed the combined mass
of the components B and C and showed that their mass is very close to
the hydrogen burning minimum mass, in good agreement with the masses of
0.080 to $0.083 M_\odot$ and 0.079 to $0.080 M_\odot$ derived by
\cite{leinert2000} from theoretical mass-luminosity relations
\citep{Baraffe1998,Chabrier2000a}. \cite{Seifahrt2008} constrained the
combined mass of B and C to $M_{B}+M_{C} = 0.157\pm0.009 M_\odot$
which is higher than \cite{Leinert2001} because of the refined
distance by \cite{Costa2005}. Recently, an improved fit for the orbit
of LHS\,1070 B and C around each other, and an estimate for the orbit
of B and C around A have been performed. The masses of the three 
components are found to be $M_A =0.13$ to $0.16 M_\odot$, 
$M_B = 0.077\pm 0.005 M_\odot$, and $M_C= 0.071\pm 0.004 M_\odot$ \citep{K2012}. 
Here, the values for the primary are uncertain, because the wide orbit
of this triple system has could not yet been determined with
sufficient accuracy.  
Finally, \cite{Seifahrt2008} have not measured individual masses for the B and C
components, but using the mass ratio $M_C/M_B=0.923$ from \citep{K2012},
their masses for A, B, C become $0.115\pm0.01$, $0.082\pm0.01$ and
$0.075\pm0.01 M_\odot$. 

The interpolations of the NextGen \cite[]{Baraffe1998} isochrone for the primary, 
and of the AMES-Dusty \cite[]{Chabrier2000a} isochrone for the B and C components,
assuming an age of 1 Gyr for the dynamical masses of \cite{Seifahrt2008} are shown in Table~\ref{tab:3}. 
For masses above the hydrogen burning limit these values do not change much for larger
ages, since in the stellar regime only show small evolution effects
are seen after the age of 1~Gyr. 
No evolution models have yet been calculated using the BT-Settl models. However,
only negligible changes are expected with the revision of the
evolution calculations. Such revision of the interior and evolution
models is currently being prepared (I.  Baraffe, Exceter, private
communication).

\subsection{Results}

\begin{figure*}[ht]
\centering
\includegraphics[width=15cm]{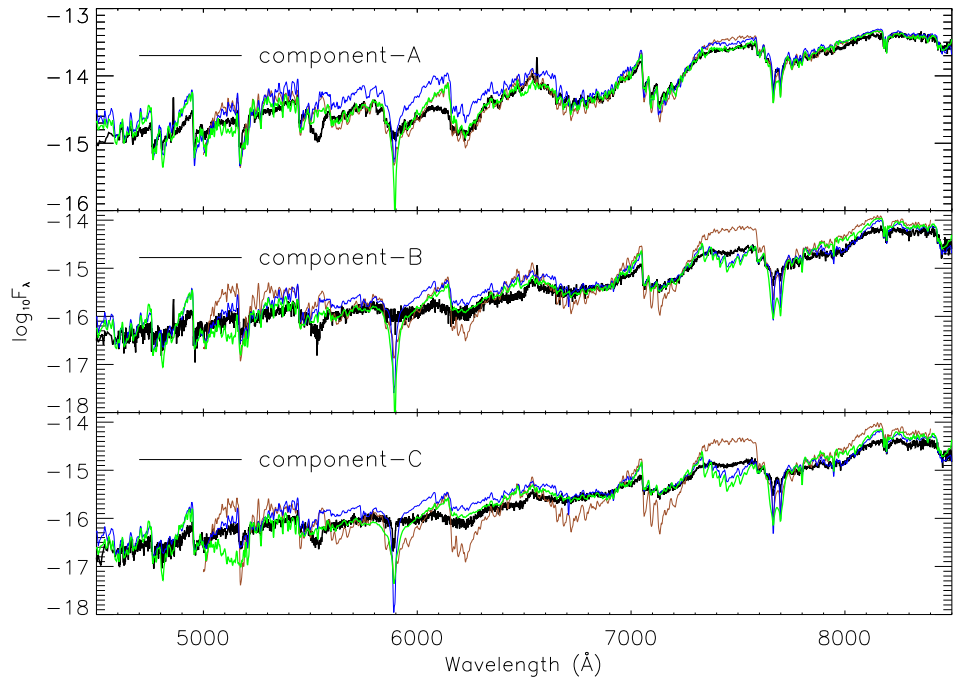}
\caption{Optical spectra of all three components. Comparison with model predictions. Black: observed FOS spectra. Green:  best fit BT-Settl model. Blue:
         best fit DRIFT model. Brown: best fit MARCS model. The parameters that give the best fit are given in Table~2.} 
          \label{Fig8}
   \end{figure*}

\begin{figure*}
\centering
\includegraphics[width=15cm]{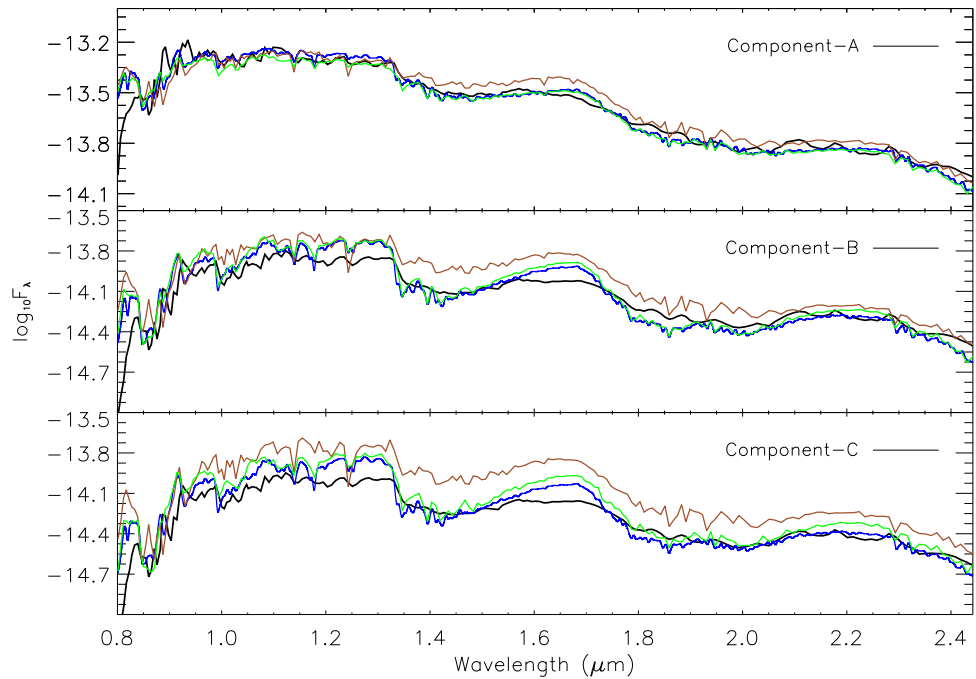}
\caption{Same as Fig.~\ref{Fig8} with NICMOS spectra in the IR.} 
          \label{Fig9}
   \end{figure*}

Figures~\ref{Fig8} and \ref{Fig9} show the best fit model superimposed to the observed optical (FOS) and IR (NICMOS) spectra for all the three components using the BT-Settl, BT-Dusty, MARCS, and DRIFT models. Note that the observed NICMOS spectra of all three components plunge down away from the model predictions below 0.85~$\mu$m (see Fig.~\ref{Fig9}), whereas the observed FOS spectra are correctly represented by the models in the same wavelength range (see Fig.~\ref{Fig8}). This deviation is due to difficulties with the NICMOS data at the very edge of the wavelength range.

The FOS spectral distribution is better reproduced by the models for the primary than for the cooler components B and C. The revised opacities and oxygen abundance (among other elements) used in the current BT-Settl models allow a significant improvement compared to the AMES-Dusty and NextGen models used in previous analysis \cite[]{Leinert1998,leinert2000}. The slope of the spectra is now reproduced over the complete FOS spectral distribution, and the strength of molecular bands is reproduced in average quite well. However some problems remain which are probably due to uncertain and missing opacity sources. Hence, the MgH feature at around 5200~$\AA$ is too strong in all the models while the CaOH band at 5500~$\AA$ is missing in all models. The NaI doublet at around 5900~$\AA$ is far too strong in the models as well as the CaH band at 7000~$\AA$. Largest discrepancies are found around 6000-6400~$\AA$ for all models. The TiO bands around 7055~$\AA$ as well as the CaH band around 6900~$\AA$ are  too strong in all the models. The VO band around 7334~$\AA$ is also visible and is quite well matched by the BT-Settl model. The BT-Settl models also differ from the DRIFT models by the strength of atomic lines which are deeper in the DRIFT models.  

\begin{figure*}[ht]
   \centering
   \includegraphics[height=6.5cm]{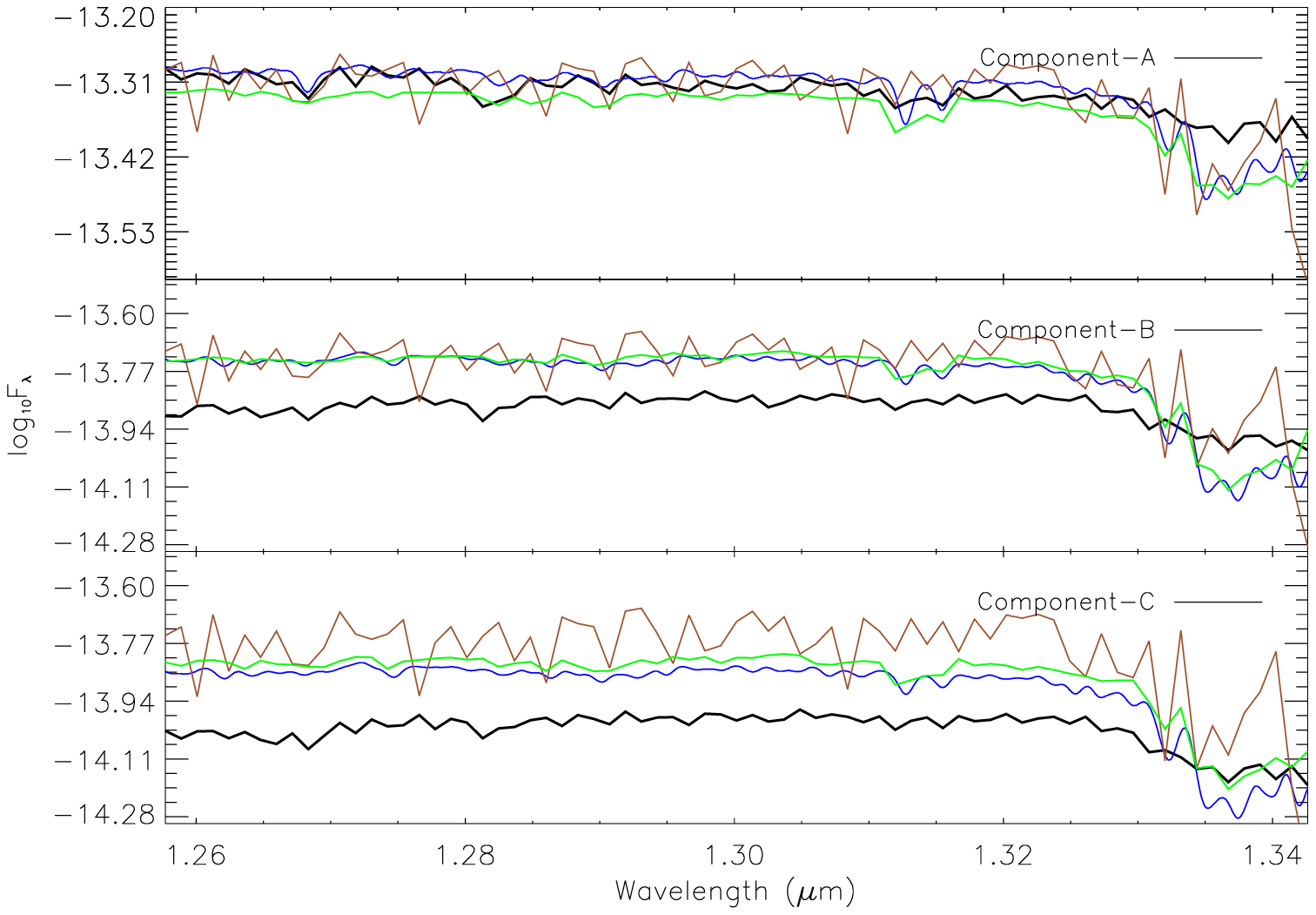}
   \includegraphics[height=6.5cm]{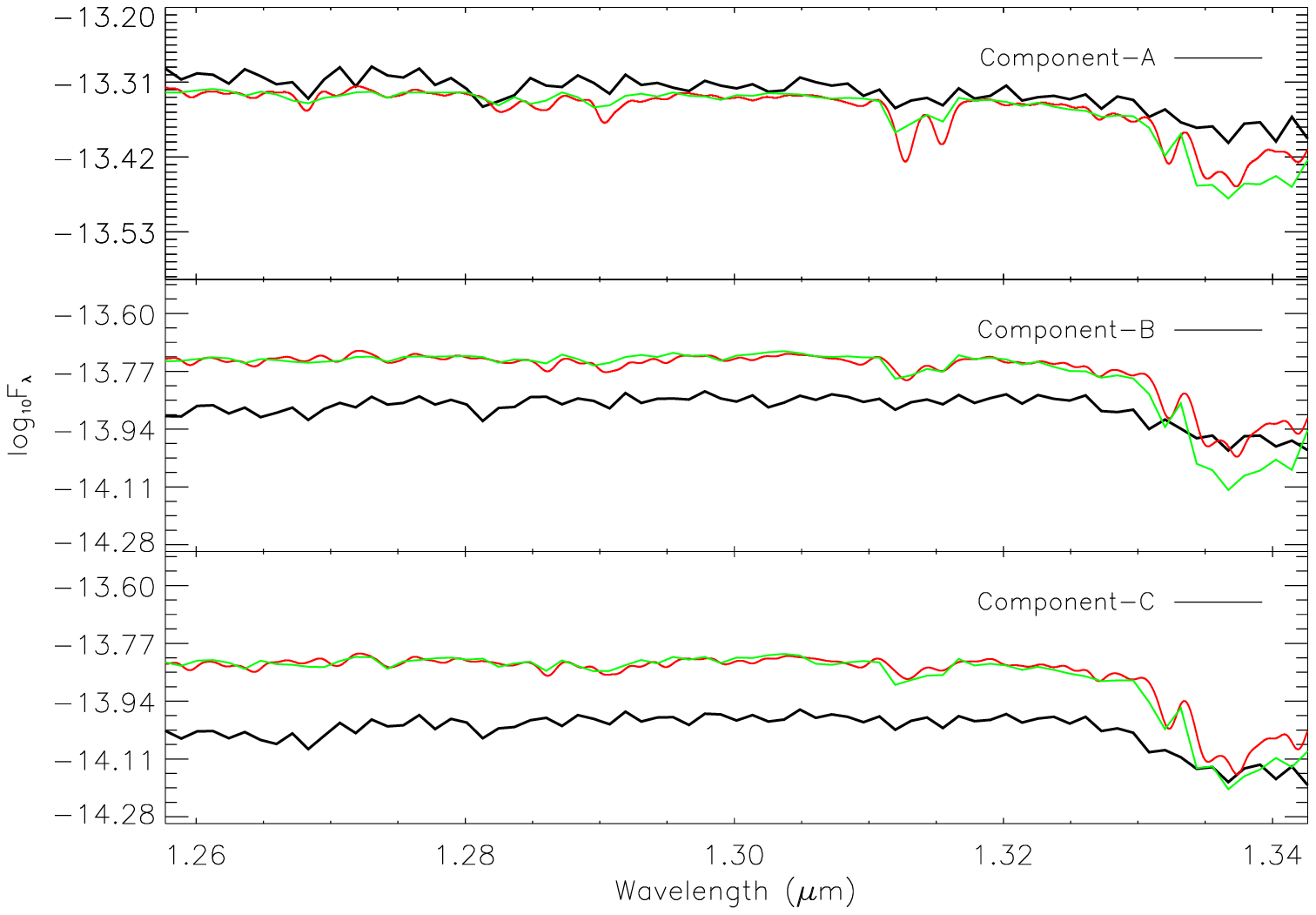}
   \includegraphics[height=6.5cm]{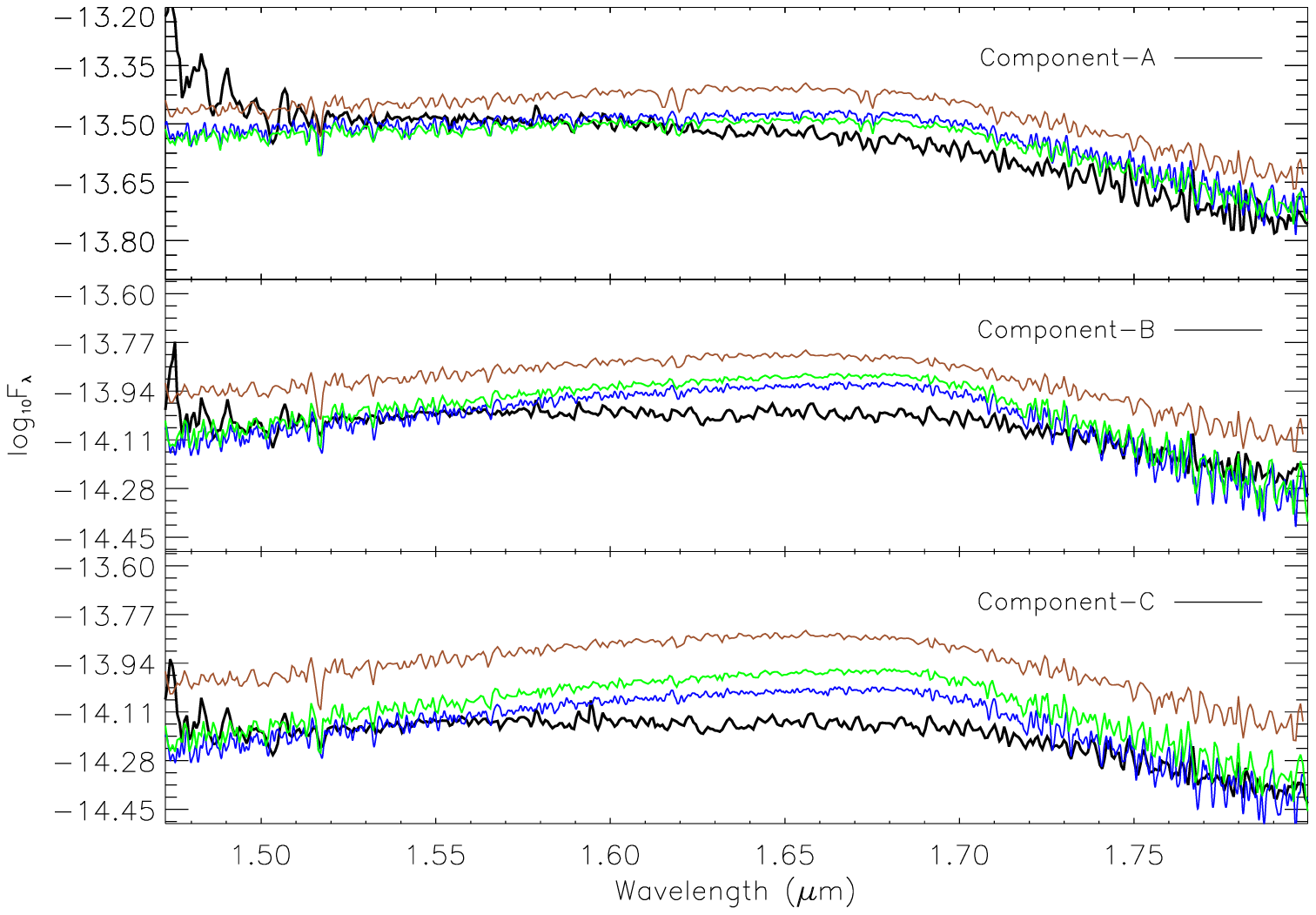}
   \includegraphics[height=6.5cm]{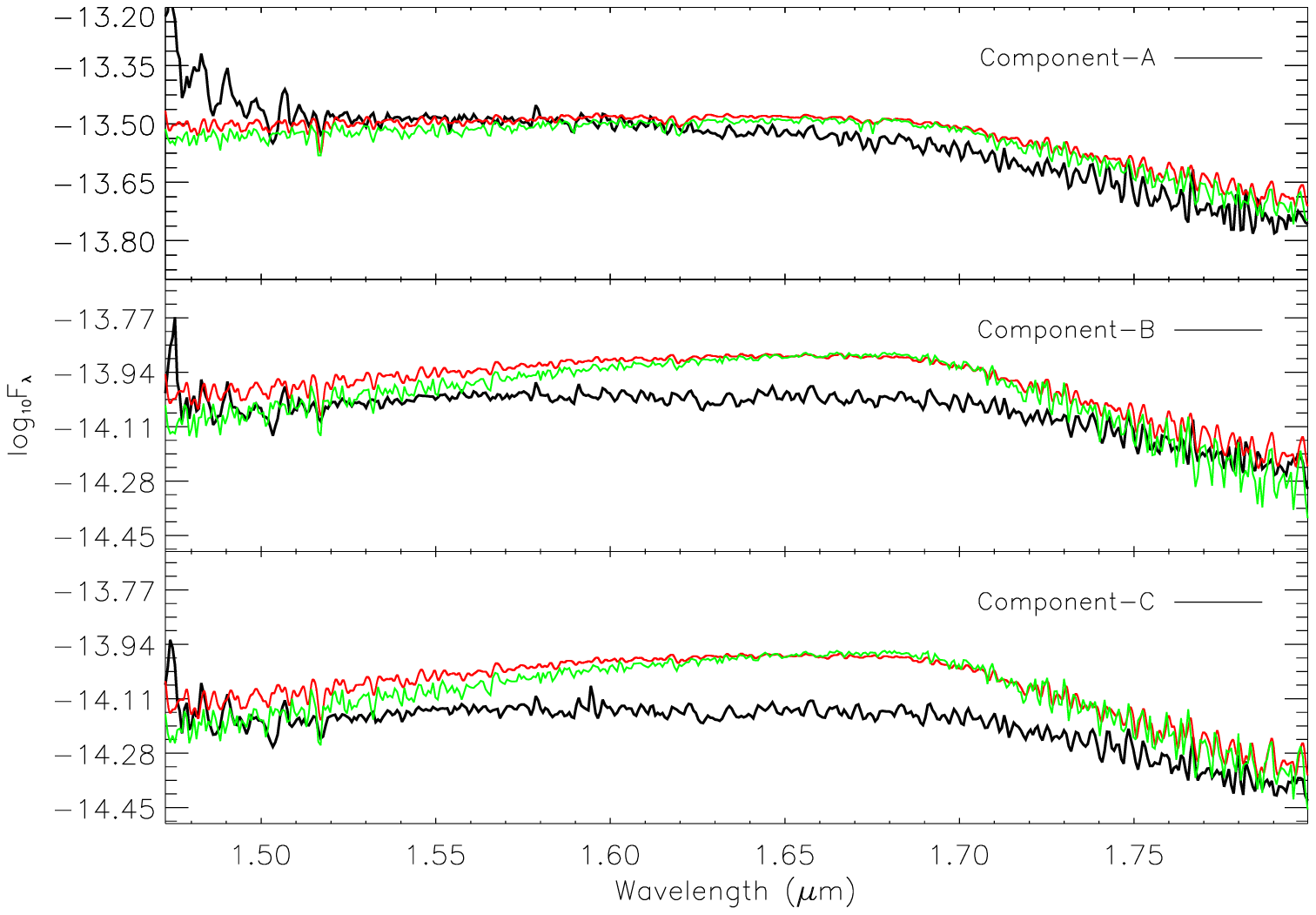}
   \includegraphics[height=6.5cm]{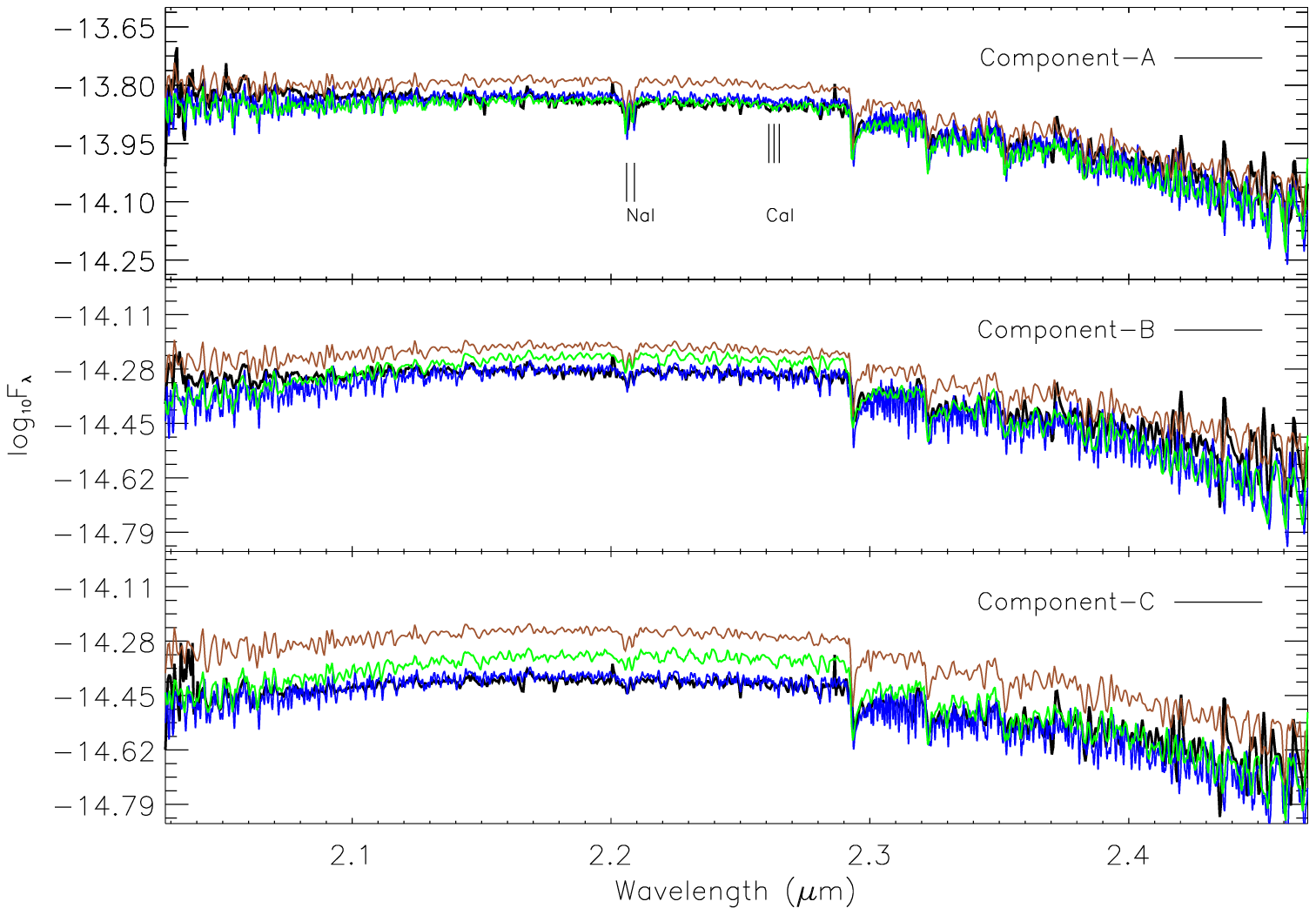}
   \includegraphics[height=6.5cm]{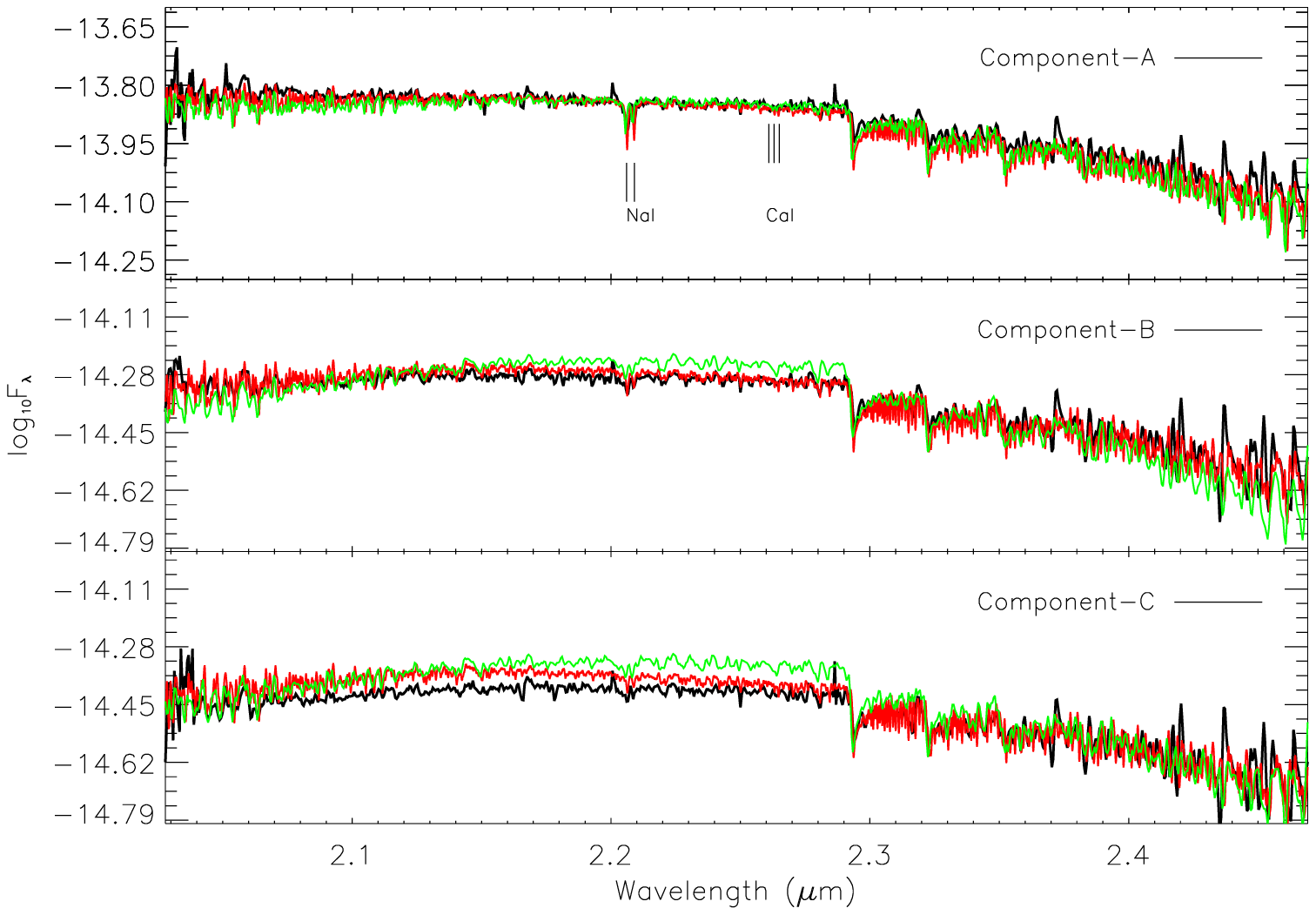}
   \caption{Black: J (upper panel), H (middle panel), and K (lower panel) NACO spectra of all three components. Comparison with model predictions. Green:  best fit BT-Settl model.  Red: best fit BT-Dusty model. Blue: best fit DRIFT model. Brown: best fit MARCS model. The comparison of the BT-Dusty and BT-Settl models together with the observations are shown on the right panels.}
         \label{Fig10}
   \end{figure*}

In the IR NICMOS range, the BT-Settl models fit slightly better the primary than the DRIFT models. The MARCS model overestimates the flux over most of the spectral range above 1.3 $\mu$m while the DRIFT model shows a slightly different shape of the H band peak which is shared by the BT-Settl models in the case of the B and C components. But all the models appear over luminous in the J and H bands in the case of the B and C components for the selected radius and effective temperature. 
This is also apparent in Figure~\ref{Fig10} which shows the comparison of high resolution IR spectra (NACO) with the best fit of all the four models. The change in the NaI and CaI strength as the temperature decreases (K-band) is quite well reproduced by the BT-Settl, BT-Dusty and DRIFT models.

\begin{figure}[ht]
\centering
\includegraphics[width=9cm]{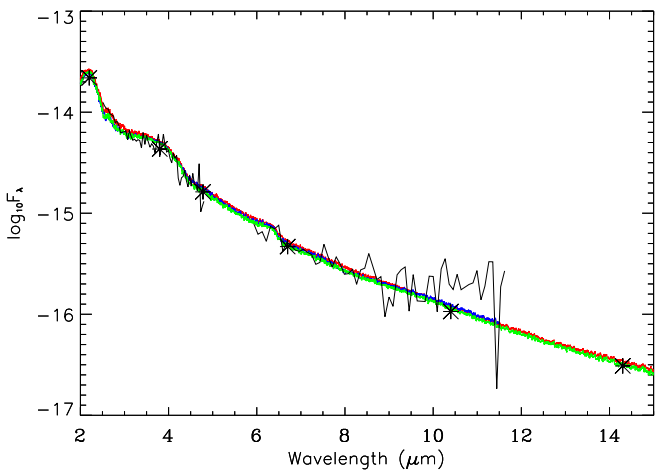}
\caption{Black: ISOPHOT thermal infrared spectra of the unresolved system, with photometric measurements 
overplotted (stars). Red: best fit BT-Dusty model. Green: best fit BT-Settl model. Blue: best fit DRIFT model} 
          \label{Fig11}
   \end{figure}

As an additional check on the effective temperature determination, we also compared the spectrum of the unresolved system in the $3-14~\mu$m range with the BT-Dusty, BT-Settl and DRIFT models (MARCS models are not available in this wavelength range). The synthetic spectra of the unresolved system are computed by adding the individual best fit synthetic spectra. The comparison is shown in Fig.~\ref{Fig11}, where the ISOPHOT spectrum is in black, the BT-Dusty model in red, the BT-Settl model in green, and the DRIFT model in blue.  Star symbols indicates the photometry obtained in the IR bands. The overall agreement is good except for the observed spectrum above $8.7 \mu$m, where it gets quite noisy.

Relatively small differences distinguish the MARCS, DRIFT and BT
models in the IR spectral range for the primary. In the case of the B
and C components, the MARCS models suffer clearly of the lack of dust
grain scattering which tends to flatten out or veil the spectral
features in this spectra range. This effect is observed in the DRIFT,
BT-Settl and BT-Dusty models which include dust formation. Differences
in the cloud model approaches explain the differences between the
DRIFT, BT-Dusty and BT-Settl models. The DRIFT models appear as dusty
as the 2001 AMES-Dusty models with similar effective temperatures and
surface gravities for B and C than derived by \cite{leinert2000}. The
BT models tends to attribute slightly hotter effective temperatures
($+$100\,K) and lower gravities ($-$0.5 dex) to these objects, while
the dust-free MARCS models would attribute them the highest values
($+$200\,K). But judging from the overall fits obtained to the NICMOS
spectra it appears that neither of the models are yet dusty enough to
explain the IR spectral distribution of B and C. Indeed the
over-luminosity of the models in the $J$ band could be attributed to
missing or to weak veiling by dust scattering. 

The \cite{K2012} mass estimate for the primary requires a 150 to 300\,K
higher effective temperature and up to 30\% larger radius than using
the revised \cite{Seifahrt2008} mass, resulting in an over-prediction
of the luminosity of up to one magnitude. 
In the case of the B and C components, the  \cite{K2012} mass estimates
correspond to effective temperatures which are 100 to 200\,K cooler 
than obtained in this paper using the DRIFT models, and in
correspondingly larger discrepancies with the BT-Settl fits. 
On the other hand, the results obtained in this paper for all three 
components are consistent with the isochrone interpolation for the 
revised \cite{Seifahrt2008} masses of the A, B and C components 
of LHS\,1070 (see Table~\ref{tab:3}), and there is no evidence that the
components may have been influenced by their binary nature. 

\section{Conclusions}
\label{ccl}

This paper presents the results from spectral synthesis analysis for
the LHS\,1070 triple system. This system has been extensively observed
from the optical to the IR, and dynamical masses have been determined
\citep{Leinert2001,Seifahrt2008}. Therefore, it constitutes a testbed
of model atmospheres of low-mass stars. Band strength indices are used
to measure TiO, CaH and PC3 features to classify their spectral
type. The components are classified as M5.5, M9.5, and L0 dwarfs, and
their atmospheres lie in a temperature range where dust starts to
form. We have
determined the physical parameters $T_\mathrm{eff}$,
$\mathrm{log}\,g$, metallicity, and radius for  the three components
of the LHS\,1070 system by comparing the observed spectra with the
synthetic spectra computed with the most recent atmospheric models:
BT-Dusty, BT-Settl, MARCS, and DRIFT. All the models agree for a solar 
metallicity for the system. The derived gravity is 5.0 dex and agrees within 
the uncertainties with the
values derived from the dynamical mass \citep{Seifahrt2008}. 

We found the same value for $T_\mathrm{eff}$ for the primary from
all models while differences of 100\,K and 200\,K are found for
components B and C depending on the dust density content of the model
atmosphere used. The revised oxygen abundance by \cite{Asplund2009}
and \cite{Caffau2011} yield significant improvements of the BT-Settl
fits to the primary compared to earlier studies based on the 
larger values of the solar oxygen abundance of \citet{Grevesse1998}. 
These improvements are described in \cite{Allard2012a}.
The even lower abundances of \citet{Grevesse2007} used in the MARCS
models lead to an excess of near-IR flux due to weaker water vapor
absorption. 
 
The DRIFT and BT-Settl models differ mainly in their numerical
approach in solving the equations for grain growth, sedimentation and
opacities: the DRIFT model solves them from the top to the bottom of the
atmosphere, while the BT-Settl model solves them from the bottom to
the top of the atmosphere. This causes the BT-Settl model to tend to
have a deficit of grains in the upper atmospheric layers compared to
the DRIFT model despite an adequate account in both models of
supersaturation effects. Despite these fundamental differences, the 
resulting grain sizes obtained by the models are quite similar.  The 
over-luminosity shared by the models in the $J$ bandpass could be 
indicative of grains of larger sizes and/or more numerous in the 
LHS1070 B and C component  atmospheres. The results confirm 
the \cite{Allard2012a} findings based on $T_\mathrm{eff}$-color constraints. 

The $T_\mathrm{eff}$ values found with the DRIFT atmospheres for components
B and C agree with the \cite{leinert2000} findings. Except for the MARCS
models which do not include dust treatment, the models are able to
reproduce the observations and describe the main features of the
visible to IR spectra for all three components. This raises the
confidence level in the dust-modelling approach. 

However, the calculation of opacities for composite grains relies on relatively
simple approximations and also does not account for possible
distributions of grain shapes and structures such as porosity. 
Both models rely on results of radiation
hydrodynamical simulations that provide the mixing and overshooting
which compensates sedimentation effects. One problem is certainly that
the translation of the resulting radial velocity field into a diffusion
coefficient is currently uncertain.  It is also possible that the
mixing effects are being currently underestimated by local 2D
simulations, and that additional mixing is provided by other phenomena
on larger scales such as global rotation effects. 

\acknowledgements
We thank Joris Blommaert for the calibration of the ISOCAM photometry, Chris Davis for obtaining the UIST data, Hongchi Wang for reducing the NACO
spectra, Tom Herbst for the MAX data, and Thomas M\"uller for calibrating them. N.R. is a Royal Swedish Academy of Sciences Research Fellow supported by a grant from the Knut and
Alice Wallenberg Foundation. N.R. acknowledges support from the Swedish Research Council, VR and funds from Kungl. Fysiografiska S\"allskapet i
Lund. N.R. thanks Dr. Kjell Eriksson for valuable help and discussions
concerning the running of the MARCS model-atmosphere program. We acknowledge financial support from "Programme National de
Physique Stellaire" (PNPS) of CNRS/INSU, France.
The research leading to these results has received funding from the ``Agence Nationale de la
Recherche'' (ANR), the ``Programme National de Physique Stellaire'' (PNPS) of CNRS (INSU), the University of Franche-Comt\'e,
and the  European Research Council under the European Community's Seventh Framework
Programme (FP7/2007-2013 Grant Agreement no. 247060).
The computations were performed at the {\sl P\^ole Scientifique de Mod\'elisation Num\'erique} (PSMN) at
the {\sl \'Ecole Normale Sup\'erieure} (ENS) in Lyon and at the {\sl Gesellschaft f{\"u}r Wissenschaftliche Datenverarbeitung
G{\"o}ttingen} in collaboration with the Institut f{\"u}r Astrophysik G{\"o}ttingen. 

\bibliographystyle{aa}
\bibliography{ref}

\begin{thebibliography}{86}
\expandafter\ifx\csname natexlab\endcsname\relax\def\natexlab#1{#1}\fi

\bibitem[{Abel {et~al.}(2011)Abel, Frommhold, Li, \& Hunt}]{Abel2011}
Abel, M., Frommhold, L., Li, X., \& Hunt, K. L.~C. 2011, The Journal of
  Physical Chemistry A, 115, 6805

\bibitem[{{Allard}(1990)}]{Allard1990}
{Allard}, F. 1990, PhD thesis, PhD thesis.~Ruprecht Karls Univ.~Heidelberg,
  (1990)

\bibitem[{{Allard} {et~al.}(1998){Allard}, {Alexander}, \&
  {Hauschildt}}]{Allard1998b}
{Allard}, F., {Alexander}, D.~R., \& {Hauschildt}, P.~H. 1998, in Astronomical
  Society of the Pacific Conference Series, Vol. 154, Cool Stars, Stellar
  Systems, and the Sun, ed. {R.~A.~Donahue \& J.~A.~Bookbinder}, 63--+

\bibitem[{{Allard} {et~al.}(2007){Allard}, {Allard}, {Homeier}, {Kielkopf},
  {McCaughrean}, \& {Spiegelman}}]{Allard2007}
{Allard}, F., {Allard}, N.~F., {Homeier}, D., {et~al.} 2007, \aap, 474, L21

\bibitem[{{Allard} {et~al.}(2003){Allard}, {Guillot}, {Ludwig}, {Hauschildt},
  {Schweitzer}, {Alexander}, \& {Ferguson}}]{Allard2003}
{Allard}, F., {Guillot}, T., {Ludwig}, H.-G., {et~al.} 2003, in IAU Symposium,
  Vol. 211, Brown Dwarfs, ed. {E.~Mart{\'{\i}}n}, 325--+

\bibitem[{{Allard} \& {Hauschildt}(1995)}]{Allard1995}
{Allard}, F. \& {Hauschildt}, P.~H. 1995, \apj, 445, 433

\bibitem[{{Allard} \& {Hauschildt}(1998)}]{Allard1998a}
{Allard}, F. \& {Hauschildt}, P.~H. 1998, in Astronomical Society of the
  Pacific Conference Series, Vol. 134, {Brown Dwarfs and Extrasolar Planets},
  ed. {{Rebolo}, R. and {Mart{\'{\i}}n}, E.~L. and {Zapatero-Osorio}, M.~R.},
  370--382

\bibitem[{{Allard} {et~al.}(1997){Allard}, {Hauschildt}, {Alexander}, \&
  {Starrfield}}]{Allard1997}
{Allard}, F., {Hauschildt}, P.~H., {Alexander}, D.~R., \& {Starrfield}, S.
  1997, \araa, 35, 137

\bibitem[{{Allard} {et~al.}(2001){Allard}, {Hauschildt}, {Alexander},
  {Tamanai}, \& {Schweitzer}}]{Allard2001}
{Allard}, F., {Hauschildt}, P.~H., {Alexander}, D.~R., {Tamanai}, A., \&
  {Schweitzer}, A. 2001, \apj, 556, 357

\bibitem[{{Allard} \& {Homeier}(2012)}]{Allard2012a}
{Allard}, F. \& {Homeier}, D. 2012, ArXiv e-prints

\bibitem[{{Asplund} {et~al.}(2009){Asplund}, {Grevesse}, {Sauval}, \&
  {Scott}}]{Asplund2009}
{Asplund}, M., {Grevesse}, N., {Sauval}, A.~J., \& {Scott}, P. 2009, \araa, 47,
  481

\bibitem[{{Baraffe} {et~al.}(1998){Baraffe}, {Chabrier}, {Allard}, \&
  {Hauschildt}}]{Baraffe1998}
{Baraffe}, I., {Chabrier}, G., {Allard}, F., \& {Hauschildt}, P.~H. 1998, \aap,
  337, 403

\bibitem[{{Barber} {et~al.}(2006){Barber}, {Tennyson}, {Harris}, \&
  {Tolchenov}}]{BT2H2O}
{Barber}, R.~J., {Tennyson}, J., {Harris}, G.~J., \& {Tolchenov}, R.~N. 2006,
  \mnras, 368, 1087

\bibitem[{{Barklem} {et~al.}(2000){Barklem}, {Piskunov}, \&
  {O'Mara}}]{Barklem2000}
{Barklem}, P.~S., {Piskunov}, N., \& {O'Mara}, B.~J. 2000, \aaps, 142, 467

\bibitem[{{Basri} \& {Marcy}(1995)}]{Basri1995}
{Basri}, G. \& {Marcy}, G.~W. 1995, \aj, 109, 762

\bibitem[{{Bonfils} {et~al.}(2011){Bonfils}, {Delfosse}, {Udry}, {Forveille},
  {Mayor}, {Perrier}, {Bouchy}, {Gillon}, {Lovis}, {Pepe}, {Queloz}, {Santos},
  {S{\'e}gransan}, \& {Bertaux}}]{Bonfils2011}
{Bonfils}, X., {Delfosse}, X., {Udry}, S., {et~al.} 2011, ArXiv e-prints

\bibitem[{Borysow {et~al.}(2001)Borysow, J{\o}rgensen, \& Fu}]{Borysow2001}
Borysow, A., J{\o}rgensen, U.~G., \& Fu, Y. 2001, Journal of Quantitative
  Spectroscopy and Radiative Transfer, 68, 235

\bibitem[{{Bosch} {et~al.}(2000){Bosch}, {Ferre-Borrull}, {Leinfellner}, \&
  {Canillas}}]{Bosch2000}
{Bosch}, S., {Ferre-Borrull}, J., {Leinfellner}, N., \& {Canillas}, A. 2000,
  Surface Science, 453, 9

\bibitem[{{Burrows} {et~al.}(2002){Burrows}, {Ram}, {Bernath}, {Sharp}, \&
  {Milsom}}]{Burrows2002}
{Burrows}, A., {Ram}, R.~S., {Bernath}, P., {Sharp}, C.~M., \& {Milsom}, J.~A.
  2002, \apj, 577, 986

\bibitem[{{Butler} {et~al.}(2004){Butler}, {Vogt}, {Marcy}, {Fischer},
  {Wright}, {Henry}, {Laughlin}, \& {Lissauer}}]{Butler2004}
{Butler}, R.~P., {Vogt}, S.~S., {Marcy}, G.~W., {et~al.} 2004, \apj, 617, 580

\bibitem[{{Caffau} {et~al.}(2011){Caffau}, {Ludwig}, {Steffen}, {Freytag}, \&
  {Bonifacio}}]{Caffau2011}
{Caffau}, E., {Ludwig}, H.-G., {Steffen}, M., {Freytag}, B., \& {Bonifacio}, P.
  2011, \solphys, 268, 255

\bibitem[{{Cesarsky} {et~al.}(1996){Cesarsky}, {Abergel}, {Agnese}, {Altieri},
  {Augueres}, {Aussel}, {Biviano}, {Blommaert}, {Bonnal}, {Bortoletto},
  {Boulade}, {Boulanger}, {Cazes}, {Cesarsky}, {Chedin}, {Claret}, {Combes},
  {Cretolle}, {Davies}, {Desert}, {Elbaz}, {Engelmann}, {Epstein},
  {Franceschini}, {Gallais}, {Gastaud}, {Gorisse}, {Guest}, {Hawarden},
  {Imbault}, {Kleczewski}, {Lacombe}, {Landriu}, {Lapegue}, {Lena}, {Longair},
  {Mandolesi}, {Metcalfe}, {Mosquet}, {Nordh}, {Okumura}, {Ott}, {Perault},
  {Perrier}, {Persi}, {Puget}, {Purkins}, {Rio}, {Robert}, {Rouan}, {Roy},
  {Saint-Pe}, {Sam Lone}, {Sargent}, {Sauvage}, {Sibille}, {Siebenmorgen},
  {Sirou}, {Soufflot}, {Starck}, {Tiphene}, {Tran}, {Ventura}, {Vigroux},
  {Vivares}, \& {Wade}}]{Cesarsky1996}
{Cesarsky}, C.~J., {Abergel}, A., {Agnese}, P., {et~al.} 1996, \aap, 315, L32

\bibitem[{{Chabrier} \& {Baraffe}(2000)}]{Chabrier2000b}
{Chabrier}, G. \& {Baraffe}, I. 2000, \araa, 38, 337

\bibitem[{{Chabrier} {et~al.}(2000){Chabrier}, {Baraffe}, {Allard}, \&
  {Hauschildt}}]{Chabrier2000a}
{Chabrier}, G., {Baraffe}, I., {Allard}, F., \& {Hauschildt}, P. 2000, \apj,
  542, 464

\bibitem[{{Cool} {et~al.}(1996){Cool}, {Piotto}, \& {King}}]{Cool1996}
{Cool}, A.~M., {Piotto}, G., \& {King}, I.~R. 1996, \apj, 468, 655

\bibitem[{{Costa} {et~al.}(2005){Costa}, {M{\'e}ndez}, {Jao}, {Henry},
  {Subasavage}, {Brown}, {Ianna}, \& {Bartlett}}]{Costa2005}
{Costa}, E., {M{\'e}ndez}, R.~A., {Jao}, W.-C., {et~al.} 2005, \aj, 130, 337

\bibitem[{{Covey} {et~al.}(2010){Covey}, {Lada}, {Rom{\'a}n-Z{\'u}{\~n}iga},
  {Muench}, {Forbrich}, \& {Ascenso}}]{Covey2010}
{Covey}, K.~R., {Lada}, C.~J., {Rom{\'a}n-Z{\'u}{\~n}iga}, C., {et~al.} 2010,
  \apj, 722, 971

\bibitem[{{Dehn} {et~al.}(2007){Dehn}, {Helling}, {Woitke}, \&
  {Hauschildt}}]{Dehn2007}
{Dehn}, M., {Helling}, C., {Woitke}, P., \& {Hauschildt}, P. 2007, in IAU
  Symposium, Vol. 239, Convection in Astrophysics, ed. {T.~Kuroda, H.~Sugama,
  R.~Kanno, \& M.~Okamoto}, 227--229

\bibitem[{{Dulick} {et~al.}(2003){Dulick}, {Bauschlicher}, {Burrows}, {Sharp},
  {Ram}, \& {Bernath}}]{Dulick2003}
{Dulick}, M., {Bauschlicher}, Jr., C.~W., {Burrows}, A., {et~al.} 2003, \apj,
  594, 651

\bibitem[{{Ferguson} {et~al.}(2005){Ferguson}, {Alexander}, {Allard}, {Barman},
  {Bodnarik}, {Hauschildt}, {Heffner-Wong}, \& {Tamanai}}]{Ferguson2005}
{Ferguson}, J.~W., {Alexander}, D.~R., {Allard}, F., {et~al.} 2005, \apj, 623,
  585

\bibitem[{{Freytag} {et~al.}(2010){Freytag}, {Allard}, {Ludwig}, {Homeier}, \&
  {Steffen}}]{Freytag2010}
{Freytag}, B., {Allard}, F., {Ludwig}, H.-G., {Homeier}, D., \& {Steffen}, M.
  2010, \aap, 513, A19+

\bibitem[{{Gabriel} {et~al.}(1997){Gabriel}, {Acosta-Pulido}, {Heinrichsen},
  {Morris}, \& {Tai}}]{Gabriel1997}
{Gabriel}, C., {Acosta-Pulido}, J., {Heinrichsen}, I., {Morris}, H., \& {Tai},
  W.-M. 1997, in Astronomical Society of the Pacific Conference Series, Vol.
  125, Astronomical Data Analysis Software and Systems VI, ed. {G.~Hunt \&
  H.~Payne}, 108

\bibitem[{{Gould} {et~al.}(1996){Gould}, {Bahcall}, \& {Flynn}}]{Gould1996}
{Gould}, A., {Bahcall}, J.~N., \& {Flynn}, C. 1996, \apj, 465, 759

\bibitem[{{Grevesse} {et~al.}(2007){Grevesse}, {Asplund}, \&
  {Sauval}}]{Grevesse2007}
{Grevesse}, N., {Asplund}, M., \& {Sauval}, A.~J. 2007, \ssr, 130, 105

\bibitem[{{Grevesse} \& {Sauval}(1998)}]{Grevesse1998}
{Grevesse}, N. \& {Sauval}, A.~J. 1998, \ssr, 85, 161

\bibitem[{{Gustafsson} {et~al.}(2008){Gustafsson}, {Edvardsson}, {Eriksson},
  {J{\o}rgensen}, {Nordlund}, \& {Plez}}]{Gustafsson2008}
{Gustafsson}, B., {Edvardsson}, B., {Eriksson}, K., {et~al.} 2008, \aap, 486,
  951

\bibitem[{{Hammersley} {et~al.}(1998){Hammersley}, {Jourdain de Muizon},
  {Kessler}, {Bouchet}, {Joseph}, {Habing}, {Salama}, \&
  {Metcalfe}}]{Hammersley.1998}
{Hammersley}, P.~L., {Jourdain de Muizon}, M., {Kessler}, M.~F., {et~al.} 1998,
  \aaps, 128, 207

\bibitem[{{Hauschildt} {et~al.}(1999){Hauschildt}, {Allard}, \&
  {Baron}}]{Hauschildt1999}
{Hauschildt}, P.~H., {Allard}, F., \& {Baron}, E. 1999, \apj, 512, 377

\bibitem[{{Hawley} {et~al.}(2002){Hawley}, {Covey}, {Knapp}, {Golimowski},
  {Fan}, {Anderson}, {Gunn}, {Harris}, {Ivezi{\'c}}, {Long}, {Lupton},
  {McGehee}, {Narayanan}, {Peng}, {Schlegel}, {Schneider}, {Spahn}, {Strauss},
  {Szkody}, {Tsvetanov}, {Walkowicz}, {Brinkmann}, {Harvanek}, {Hennessy},
  {Kleinman}, {Krzesinski}, {Long}, {Neilsen}, {Newman}, {Nitta}, {Snedden}, \&
  {York}}]{Hawley2002}
{Hawley}, S.~L., {Covey}, K.~R., {Knapp}, G.~R., {et~al.} 2002, \aj, 123, 3409

\bibitem[{{Helling} {et~al.}(2008{\natexlab{a}}){Helling}, {Ackerman},
  {Allard}, {Dehn}, {Hauschildt}, {Homeier}, {Lodders}, {Marley}, {Rietmeijer},
  {Tsuji}, \& {Woitke}}]{Helling2008c}
{Helling}, C., {Ackerman}, A., {Allard}, F., {et~al.} 2008{\natexlab{a}},
  \mnras, 391, 1854

\bibitem[{{Helling} {et~al.}(2008{\natexlab{b}}){Helling}, {Dehn}, {Woitke}, \&
  {Hauschildt}}]{Helling2008b}
{Helling}, C., {Dehn}, M., {Woitke}, P., \& {Hauschildt}, P.~H.
  2008{\natexlab{b}}, \apjl, 675, L105

\bibitem[{{Helling} {et~al.}(2008{\natexlab{c}}){Helling}, {Woitke}, \&
  {Thi}}]{Helling2008a}
{Helling}, C., {Woitke}, P., \& {Thi}, W.-F. 2008{\natexlab{c}}, \aap, 485, 547

\bibitem[{{Henry}(1998)}]{Henry1998}
{Henry}, T.~J. 1998, in Astronomical Society of the Pacific Conference Series,
  Vol. 134, Brown Dwarfs and Extrasolar Planets, ed. {R.~Rebolo, E.~L.~Martin,
  \& M.~R.~Zapatero Osorio}, 28--+

\bibitem[{{Henry} {et~al.}(1999){Henry}, {Franz}, {Wasserman}, {Benedict},
  {Shelus}, {Ianna}, {Kirkpatrick}, \& {McCarthy}}]{Henry1999}
{Henry}, T.~J., {Franz}, O.~G., {Wasserman}, L.~H., {et~al.} 1999, \apj, 512,
  864

\bibitem[{{Homeier}(2005)}]{Homeier2005}
{Homeier}, D. 2005, Memorie della Societ{\`a} Astronomica Italiana Supplementi,
  7, 157

\bibitem[{{Homeier} {et~al.}(2003){Homeier}, {Hauschildt}, \&
  {Allard}}]{Homeier2003}
{Homeier}, D., {Hauschildt}, P., \& {Allard}, F. 2003, in ASP Conference
  Series, Vol. 288, Stellar Atmosphere Modeling, ed. I.~Hubeny, D.~Mihalas, \&
  K.~Werner (San Francisco: Astronomical Society of the Pacific), 357--360

\bibitem[{{K{\"o}hler} {et~al.}(2012){K{\"o}hler}, {Ratzka}, \&
  {Leinert}}]{K2012}
{K{\"o}hler}, R., {Ratzka}, T., \& {Leinert}, C. 2012, \aap, 541, A29

\bibitem[{{Kupka} \& {Ryabchikova}(1999)}]{Kupka1999}
{Kupka}, F. \& {Ryabchikova}, T.~A. 1999, Publications de l'Observatoire
  Astronomique de Beograd, 65, 223

\bibitem[{{Leggett} {et~al.}(1998){Leggett}, {Allard}, \&
  {Hauschildt}}]{Leggett1998}
{Leggett}, S.~K., {Allard}, F., \& {Hauschildt}, P.~H. 1998, \apj, 509, 836

\bibitem[{{Leggett} {et~al.}(1994){Leggett}, {Harris}, \& {Dahn}}]{Leggett1994}
{Leggett}, S.~K., {Harris}, H.~C., \& {Dahn}, C.~C. 1994, \aj, 108, 944

\bibitem[{{Leinert} {et~al.}(2000){Leinert}, {Allard}, {Richichi}, \&
  {Hauschildt}}]{leinert2000}
{Leinert}, C., {Allard}, F., {Richichi}, A., \& {Hauschildt}, P.~H. 2000, \aap,
  353, 691

\bibitem[{{Leinert} {et~al.}(2001){Leinert}, {Jahrei{\ss}}, {Woitas}, {Zucker},
  {Mazeh}, {Eckart}, \& {K{\"o}hler}}]{Leinert2001}
{Leinert}, C., {Jahrei{\ss}}, H., {Woitas}, J., {et~al.} 2001, \aap, 367, 183

\bibitem[{{Leinert} {et~al.}(1994){Leinert}, {Weitzel}, {Richichi}, {Eckart},
  \& {Tacconi-Garman}}]{Leinert1994}
{Leinert}, C., {Weitzel}, N., {Richichi}, A., {Eckart}, A., \&
  {Tacconi-Garman}, L.~E. 1994, \aap, 291, L47

\bibitem[{{Leinert} {et~al.}(1998){Leinert}, {Woitas}, {Allard}, {Richichi}, \&
  {Jahreiss}}]{Leinert1998}
{Leinert}, C., {Woitas}, J., {Allard}, F., {Richichi}, A., \& {Jahreiss}, H.
  1998, in Astronomical Society of the Pacific Conference Series, Vol. 134,
  Brown Dwarfs and Extrasolar Planets, ed. {R.~Rebolo, E.~L.~Martin, \&
  M.~R.~Zapatero Osorio}, 203

\bibitem[{{Lemke} {et~al.}(1996){Lemke}, {Klaas}, {Abolins}, {Abraham},
  {Acosta-Pulido}, {Bogun}, {Castaneda}, {Cornwall}, {Drury}, {Gabriel},
  {Garzon}, {Gemuend}, {Groezinger}, {Gruen}, {Haas}, {Hajduk}, {Hall},
  {Heinrichsen}, {Herbstmeier}, {Hirth}, {Joseph}, {Kinkel}, {Kirches},
  {Koempe}, {Kraetschmer}, {Kreysa}, {Krueger}, {Kunkel}, {Laureijs},
  {Luetzow-Wentzky}, {Mattila}, {Mueller}, {Pacher}, {Pelz}, {Popow},
  {Rasmussen}, {Rodriguez Espinosa}, {Richards}, {Russell}, {Schnopper},
  {Schubert}, {Schulz}, {Telesco}, {Tilgner}, {Tuffs}, {Voelk}, {Walker},
  {Wells}, \& {Wolf}}]{Lemke1996}
{Lemke}, D., {Klaas}, U., {Abolins}, J., {et~al.} 1996, \aap, 315, L64

\bibitem[{{Lenzen} {et~al.}(2003){Lenzen}, {Hartung}, {Brandner}, {Finger},
  {Hubin}, {Lacombe}, {Lagrange}, {Lehnert}, {Moorwood}, \&
  {Mouillet}}]{Lenzen2003}
{Lenzen}, R., {Hartung}, M., {Brandner}, W., {et~al.} 2003, Proc. SPIE, 4841,
  944

\bibitem[{{L{\'e}pine} {et~al.}(2003){L{\'e}pine}, {Rich}, \&
  {Shara}}]{Lepine2003a}
{L{\'e}pine}, S., {Rich}, R.~M., \& {Shara}, M.~M. 2003, \aj, 125, 1598

\bibitem[{{Lodders} \& {Fegley}(2006)}]{Lodders2006}
{Lodders}, K. \& {Fegley}, Jr., B. 2006, {Chemistry of Low Mass Substellar
  Objects}, ed. {Mason, J.~W.} (Springer Verlag), 1--+

\bibitem[{{Ludwig} {et~al.}(2002){Ludwig}, {Allard}, \&
  {Hauschildt}}]{Ludwig2002}
{Ludwig}, H.-G., {Allard}, F., \& {Hauschildt}, P.~H. 2002, \aap, 395, 99

\bibitem[{{Ludwig} {et~al.}(2006){Ludwig}, {Allard}, \&
  {Hauschildt}}]{Ludwig2006}
{Ludwig}, H.-G., {Allard}, F., \& {Hauschildt}, P.~H. 2006, \aap, 459, 599

\bibitem[{{Ludwig} {et~al.}(1999){Ludwig}, {Freytag}, \&
  {Steffen}}]{Ludwig1999}
{Ludwig}, H.-G., {Freytag}, B., \& {Steffen}, M. 1999, \aap, 346, 111

\bibitem[{{Maiolino} {et~al.}(1996){Maiolino}, {Rieke}, \&
  {Rieke}}]{Maiolino1996}
{Maiolino}, R., {Rieke}, G.~H., \& {Rieke}, M.~J. 1996, \aj, 111, 537

\bibitem[{{Mart{\'{\i}}n} {et~al.}(1999){Mart{\'{\i}}n}, {Delfosse}, {Basri},
  {Goldman}, {Forveille}, \& {Zapatero Osorio}}]{Martin1999}
{Mart{\'{\i}}n}, E.~L., {Delfosse}, X., {Basri}, G., {et~al.} 1999, \aj, 118,
  2466

\bibitem[{{Mera} {et~al.}(1996){Mera}, {Chabrier}, \& {Baraffe}}]{Mera1996}
{Mera}, D., {Chabrier}, G., \& {Baraffe}, I. 1996, \apjl, 459, L87+

\bibitem[{{Plez}(1998)}]{Plez1998}
{Plez}, B. 1998, \aap, 337, 495

\bibitem[{{Reid} {et~al.}(1995){Reid}, {Hawley}, \& {Gizis}}]{Reid1995}
{Reid}, I.~N., {Hawley}, S.~L., \& {Gizis}, J.~E. 1995, \aj, 110, 1838

\bibitem[{{Reid}(1993)}]{Reid1993}
{Reid}, N. 1993, \mnras, 265, 785

\bibitem[{{Reiners} {et~al.}(2007{\natexlab{a}}){Reiners}, {Homeier},
  {Hauschildt}, \& {Allard}}]{Reiners2007b}
{Reiners}, A., {Homeier}, D., {Hauschildt}, P.~H., \& {Allard}, F.
  2007{\natexlab{a}}, \aap, 473, 245

\bibitem[{{Reiners} {et~al.}(2007{\natexlab{b}}){Reiners}, {Seifahrt},
  {K{\"a}ufl}, {Siebenmorgen}, \& {Smette}}]{Reiners2007a}
{Reiners}, A., {Seifahrt}, A., {K{\"a}ufl}, H.~U., {Siebenmorgen}, R., \&
  {Smette}, A. 2007{\natexlab{b}}, \aap, 471, L5

\bibitem[{{Renzini} {et~al.}(1996){Renzini}, {Bragaglia}, {Ferraro},
  {Gilmozzi}, {Ortolani}, {Holberg}, {Liebert}, {Wesemael}, \&
  {Bohlin}}]{Renzini1996}
{Renzini}, A., {Bragaglia}, A., {Ferraro}, F.~R., {et~al.} 1996, \apjl, 465,
  L23+

\bibitem[{{Robberto} \& {Herbst}(1998)}]{Robberto1998}
{Robberto}, M. \& {Herbst}, T.~M. 1998, Proc. SPIE, 3354, 711

\bibitem[{{Rojas-Ayala} {et~al.}(2010){Rojas-Ayala}, {Covey}, {Muirhead}, \&
  {Lloyd}}]{Rojas-Ayala2010}
{Rojas-Ayala}, B., {Covey}, K.~R., {Muirhead}, P.~S., \& {Lloyd}, J.~P. 2010,
  \apjl, 720, L113

\bibitem[{{Rossow}(1978)}]{Rossow78}
{Rossow}, W.~B. 1978, \icarus, 36, 1

\bibitem[{{Rousset} {et~al.}(2003){Rousset}, {Lacombe}, {Puget}, {Hubin},
  {Gendron}, {Fusco}, {Arsenault}, {Charton}, {Feautrier}, {Gigan}, {Kern},
  {Lagrange}, {Madec}, {Mouillet}, {Rabaud}, {Rabou}, {Stadler}, \&
  {Zins}}]{Rousset2003}
{Rousset}, G., {Lacombe}, F., {Puget}, P., {et~al.} 2003, Proc. SPIE, 4839, 140

\bibitem[{{Ruiz} {et~al.}(1997){Ruiz}, {Leggett}, \& {Allard}}]{Ruiz1997}
{Ruiz}, M.~T., {Leggett}, S.~K., \& {Allard}, F. 1997, \apjl, 491, L107

\bibitem[{{Seifahrt} {et~al.}(2008){Seifahrt}, {R{\"o}ll}, {Neuh{\"a}user},
  {Reiners}, {Kerber}, {K{\"a}ufl}, {Siebenmorgen}, \& {Smette}}]{Seifahrt2008}
{Seifahrt}, A., {R{\"o}ll}, T., {Neuh{\"a}user}, R., {et~al.} 2008, \aap, 484,
  429

\bibitem[{{Skory} {et~al.}(2003){Skory}, {Weck}, {Stancil}, \&
  {Kirby}}]{Skory2003}
{Skory}, S., {Weck}, P.~F., {Stancil}, P.~C., \& {Kirby}, K. 2003, \apjs, 148,
  599

\bibitem[{{Skumanich}(1972)}]{Skumanich1972}
{Skumanich}, A. 1972, \apj, 171, 565

\bibitem[{{Tashkun} {et~al.}(2004){Tashkun}, {Perevalov}, {Teffo}, {Bykov},
  {Lavrentieva}, \& {Babikov}}]{Tashkun2004}
{Tashkun}, S.~A., {Perevalov}, V.~I., {Teffo}, J.-L., {et~al.} 2004, Proc.
  SPIE, 5311, 102

\bibitem[{{Tsuji} {et~al.}(1996{\natexlab{a}}){Tsuji}, {Ohnaka}, \&
  {Aoki}}]{Tsuji1996a}
{Tsuji}, T., {Ohnaka}, K., \& {Aoki}, W. 1996{\natexlab{a}}, \aap, 305, L1+

\bibitem[{{Tsuji} {et~al.}(1999){Tsuji}, {Ohnaka}, \& {Aoki}}]{Tsuji1999}
{Tsuji}, T., {Ohnaka}, K., \& {Aoki}, W. 1999, \apjl, 520, L119

\bibitem[{{Tsuji} {et~al.}(1996{\natexlab{b}}){Tsuji}, {Ohnaka}, {Aoki}, \&
  {Nakajima}}]{Tsuji1996b}
{Tsuji}, T., {Ohnaka}, K., {Aoki}, W., \& {Nakajima}, T. 1996{\natexlab{b}},
  \aap, 308, L29

\bibitem[{Uns{\"o}ld(1968)}]{unsoeld55}
Uns{\"o}ld, A. 1968, Physik der Sternatmosph{\"a}ren, 2nd edn. (Heidelberg:
  Springer Verlag)

\bibitem[{{Valenti} \& {Piskunov}(1996)}]{Valenti1996}
{Valenti}, J.~A. \& {Piskunov}, N. 1996, \aaps, 118, 595

\bibitem[{{Witte} {et~al.}(2009){Witte}, {Helling}, \&
  {Hauschildt}}]{Witte2009}
{Witte}, S., {Helling}, C., \& {Hauschildt}, P.~H. 2009, \aap, 506, 1367

\bibitem[{{Woitke} \& {Helling}(2004)}]{Woitke2004}
{Woitke}, P. \& {Helling}, C. 2004, \aap, 414, 335

\end{thebibliography}
\end{document}